\documentclass[twocolumn,authoryear]{aastex63}
\usepackage{appendix}

\usepackage{graphicx}	% Including figure files
\usepackage{amsmath}	% Advanced maths commands
\usepackage{amssymb}	% Extra maths symbols
\usepackage{natbib}
\usepackage{times}
\usepackage{multirow}
\usepackage{xcolor}

\usepackage{url}
\usepackage{etoolbox}

\setcitestyle{super,comma}
\usepackage{hyperref}
\shortauthors{Maiti et al.}

\begin{document}
%\title{Numerical studies of plasma
%bursts due to Alfv\'en wave - energetic particle %interaction in the inner Van Allen radiation %%belt: seismic event detection and predictions %for IITM nano-satellite mission}

\title{Numerical Study of Alfv\'en Wave-Energetic Particle Interaction in the Inner Van Allen Belt and predictions of Seismic-Related \textcolor{black}{Energetic Proton} Bursts for the IITMSAT Mission}

\correspondingauthor{Snehanshu Maiti(*), snehanshu.maiti@gmail.com}

\author[0000-0002-0786-7307]{Snehanshu Maiti(*)}
\affiliation{Department of Electrical Engineering,  \\
Indian Institute of Technology Madras \\
 Chennai, 600036, India}

\author{Harishankar Ramachandran
}
\affiliation{Department of Electrical Engineering,  \\
Indian Institute of Technology Madras \\
 Chennai, 600036, India}

\begin{abstract}

The IIT Madras nano-satellite aims to investigate the science of energetic particle precipitation from the inner Van Allen (VA) radiation belt into the upper ionosphere as a potential precursor to earthquakes. Precursors in the form of low-frequency electromagnetic waves can appear several hours before an earthquake. These waves, captured near the ionosphere-magnetosphere transition region, propagate along geomagnetic field lines as Alfv\'en waves and interact resonantly with trapped energetic particles in the radiation belt, causing their precipitation. Such precipitation can be observed by satellites as \textcolor{black}{energetic} particle bursts occurring a few hours prior to the earthquake. A numerical study of Alfv\'en wave - energetic proton interactions in the inner Van Allen belt is presented here to investigate the  energetic proton precipitation and make predictions to support the scientific objective of the IITM satellite mission. A kinetic model of the energetic trapped proton population in the inner belt is developed, yielding a steady-state distribution that reproduces the observed density profile.
The Finite-Difference Time-Domain (FDTD) method is employed to simulate both narrowband seismic event–specific emissions and broadband background noise representing magnetohydrodynamic (MHD) Alfvén wave activity in the inner radiation belt. The studies of interactions of narrow-band Alfv\'en wave packets with the energetic protons in the belt reveals that a sharp cyclotron resonance condition arises at a low Alfv\'en frequency (10 Hz), causing substantial \textcolor{black}{precipitation} of high energy protons (125 MeV)   from their stable mirror orbits. This precipitation can be clearly distinguished from background noisy interactions. Based on these results, we predict the optimal satellite orbital altitude for detecting such \textcolor{black}{energetic proton} bursts.

\end{abstract}


\section{Introduction}

High energy charged particle bursts, which are sharp, short-time increase in energetic particle flux, have been observed by several satellites in  near-Earth space and reported over the past four decades \citep{Voronov1987,Voronov1989,Galperin92,Aleshina92,Galper95,Sgrigna2005,Bakhaldin2007,Aleksandrin2009,Fidani2010,Bocchini2013,Zhang2014,Aleksandrin2015, Zharaspayev2017, Fidani2018,Wang2023}. These energetic particle bursts are thought to originate from various sources, including solar–magnetospheric processes \citep{Aleksandrin2009,Zhang2021}, geophysical events, or man-made electromagnetic emissions. Among geophysical sources, such bursts could be caused by lightning or seismic activity. Our current study focuses specifically on investigating energetic particle bursts associated with seismic events and their potential as precursors to earthquakes. The correlation between earthquakes and energetic particle bursts was first reported based on observations from the MARIA experiment onboard the Salut-7 spacecraft in 1985 \citep{Voronov1987}. Since then, several other satellite missions, including MARIA-2 and SAMPLEX \citep{Sgrigna2005} at high energies (5–50 MeV), POES\citep{Fidani2008,Fidani2010,Bocchini2013,Fidani2018} and DEMETER \citep{16,18,Li2012,Zhang2013,Zhang2014} at lower energies (0.3–5 MeV),  ARINA \citep{Bakhaldin2007,Aleksandrin2015,Zharaspayev2017} and VSPLESK (3–100 MeV), and more recently CSES \citep{Wang2023} have reported similar correlations. Most of these observations involve energetic electrons with energies on the order of tens of MeV.

The inner Van Allen belt consists of very high-energy protons with energies exceeding 100~MeV, and high-energy electrons in the range of several hundred~\textcolor{black}{keV} \citep{1}. These energetic particles, which originate primarily from the solar wind and cosmic rays, are trapped by the Earth's magnetic field, forming the radiation belts. The mechanism by which energetic trapped particles in the inner radiation belt are lost due to local disturbances caused by pre-seismic events, leading to \textcolor{black}{energetic particle bursts}, is believed to be as follows.
 
 It has been proposed \cite{Galperin92,Aleshina92,Galper95,Krechtov96} that pre-seismic Ultra Low Frequency (ULF) (f $<$ 5 Hz) and Extra Low Frequency (ELF) (f = 30 - 300 Hz)  waves  are generated  by direct or indirect means several hours before the main shock \citep{Galper99,Hattori2004,Hayakawa2007,Athanou2011,Zhima2012,Winata2020,Alimoradi2024}  in a seismic event. These electromagnetic waves generated during the development of seismic activities, travel from the hypocenter region through the intervening ground, the atmosphere and reach the ionosphere. During their propagation through the solid crust, the higher frequency content of the  Seismo-Electro-Magnetic Emissions is severely attenuated. On the other hand, the ULF and ELF waves with frequencies less than or around 10 Hz  suffer the least attenuation through ground and atmosphere before they reach the upper ionosphere \citep{11a}.  They are captured near the ionosphere-magnetosphere transition region \citep{Molchanov92} and from this region, these pre-seismic ULF/ELF waves  travel along the geomagnetic field lines as Alfv\'en waves which is a kind of Magnetohydrodynamic (MHD) wave and reach the inner boundary of the inner Van-Allen radiation belt. These waves are low frequency transverse electromagnetic waves  where ions oscillate in response to the restoring force provided by the tension in the magnetic field lines. Alfv\'en waves could interact resonantly with the trapped energetic particles near the inner Van-Allen radiation belt boundary and cause their precipitation due to pitch angle diffusion \citep{16a}. The  pitch angle diffusion changes the \textcolor{black}{energetic} trapped particle's pitch angle (the angle between the particle velocity vector and magnetic field vector) and results in a decrease of their mirror point altitude (in comparison with stable trapped energetic particles). The \textcolor{black}{energetic} particles precipitate from the inner Van Allen radiation belt to the upper ionosphere, while in the same L-shell and are observed as \textcolor{black}{energetic} particle bursts (sudden increase in particle count rates) a few hours before the manifestation of an earthquake  by satellites in low-earth orbits.
 
The IITM nano-satellite mission \citep{19} is designed to study energetic particle bursts associated with seismic events. The present work aims to support the scientific objectives of the IITM mission by investigating the mechanisms of energetic particle precipitation as a potential ionospheric precursor to earthquakes using numerical simulations. \textcolor{black}{The IITM nano-satellite is primarily based on the Space-Based Proton Electron Energy Detector (SPEED) payload, which is a plastic scintillator–based charged-particle detector capable of measuring the energy spectrum of protons and electrons with energies in the ranges of 17–100 MeV and 1–15 MeV, respectively. The dimensions of the payload are $270 \times 269 \times 88$~mm$^3$, with a mass of 7~kg and an average power consumption of 3~W. It has a large active area of $225 \times 225$~mm$^2$ and a wide field of view of 178.5$^\circ$, allowing it to measure particle burst events of relatively low flux. The energy resolution of the detector is 5~MeV for protons and 1~MeV for electrons, with a high temporal resolution of 0.1~s, sufficient to characterize the time profile of particle bursts. The detector is aligned with a 0$^\circ$ pitch angle with respect to the magnetic field line, i.e., its opening face points along the magnetic field to capture low-pitch-angle precipitating particles. The satellite is planned to orbit at altitudes around 600--800~km.}

We start our study with the examination of possibilities of \textcolor{black} {energetic} proton bursts due to wave-particle interaction in the inner VA radiation belt at an L shell of 1.5. Our focus is to look for resonance conditions between \textcolor{black}{energetic} protons and Alfv\'en waves in this study. 
We further examine whether the dominant resonance mechanism corresponds to cyclotron or bounce resonance. Cyclotron resonance takes place when the Alfv\'en wave frequency is Doppler shifted to the cyclotron frequency of particle in radiation belt. Bounce resonance occurs when the wave frequency matches the particle bounce frequency (or its harmonics) along the magnetic field line, enabling resonant interaction over successive mirror-to-mirror motions. \textcolor{black}{The present work does not model the generation of Alfvénic disturbances from a realistic seismic source, nor does it provide a self-consistent lithosphere–atmosphere–ionosphere–magnetosphere transmission calculation. The present model also does not explicitly account for the transmission and damping of seismo-electromagnetic (SEME) ULF/ELF waves through the ionosphere. Instead, it assumes that only the low-frequency component of the source SEME spectrum that survives attenuation from the hypocenter reaches the ionosphere–magnetosphere transition region, from where it is treated as an Alfvénic disturbance propagating along the geomagnetic field line. The purpose of the present work is therefore not to establish the earthquake coupling chain from first principles, but rather to examine the wave–particle interaction consequences in the inner belt if weak low-frequency disturbances generated by pre-seismic events are present there. Within this reduced framework, the present study focuses on the subsequent wave–particle interaction and resulting proton precipitation in the inner radiation belt after the MHD disturbance is established along the field line, rather than modeling the full atmosphere–ionosphere transmission problem.}

The manuscript is organized as follows. In Section 2, we present a kinetic model of trapped energetic protons in the inner Van Allen radiation belt using analytical methods and numerical simulations.
In our previous work\citep{32}, we presented a first kinetic model of energetic particles in the inner radiation belt that reproduced the observed particle density distribution in the belt. Building on that foundation, in this paper we develop an improved kinetic model in Section 2, which treats energy and magnetic moment as independent variables, thereby enabling a more realistic two-dimensional representation of the $E–\mu$ phase-space distribution. We present results from numerical simulations of MHD Alfv'en wave packets in the inner radiation belt using the Finite-Difference Time-Domain (FDTD) method in Section 3. We consider two distinct types of wave inputs to represent different characteristics of seismo-electromagnetic emissions. Seismic emissions are often observed as narrowband enhancements superposed on broadband background noise. Accordingly, a narrowband coherent wave packet centered at selected ULF frequencies is used to model event-specific emissions and to examine whether they produce strong resonance-driven precipitation. In addition, a broadband incoherent spectrum, implemented as white noise with a flat power spectral density ($P(f)=\mathrm{constant}$), is included as a simplified, idealized stochastic representation of background emissions. This background is introduced to examine how its interaction with energetic particles compares with that of event-specific emissions, and whether it produces comparable precipitation or resonance-driven signatures. We present in Section 4 the results of the interaction of Alfv'en wave packets (representing seismic emissions and background noise) with steady-state trapped energetic \textcolor{black}{protons} in the inner radiation belt, examining the occurrence of resonance-driven interactions due to ULF seismo-electromagnetic waves and making predictions for the optimal satellite altitude for IITMSAT to detect energetic proton bursts. The results of the present study and the corresponding predictions for \textcolor{black}{energetic} particle burst detection for IITMSAT are summarized in Section 5.

\section{A Kinetic model of the inner Van Allen radiation belt}

\begin{figure}[h]
\centering
  \includegraphics[width=0.8\linewidth]{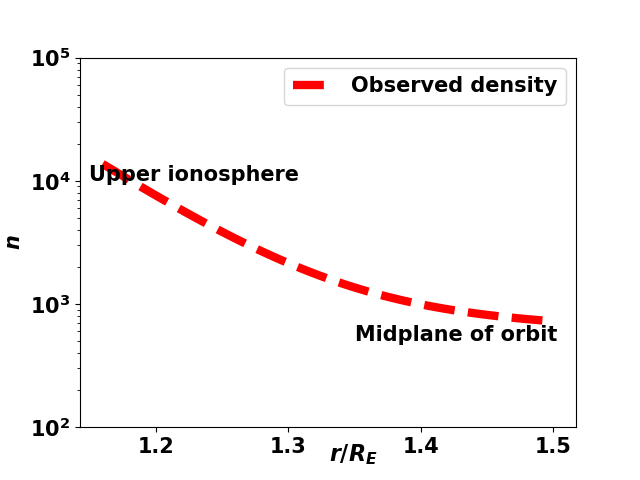}
  \vspace{1em}
 \caption{Observed density profile of the radiation belt and its variation with altitude at the equator. The $y$-axis represents the density of particles in $/cm^3$ whereas the $x$-axis represents the geocentric radius of Earth normalised by the Earth's radius or the L shell value.}
  \label{fig:observed_density}
\end{figure}

The inner Van Allen radiation belt is a region of energetic charged particles located in the lower terrestrial magnetosphere, extending from approximately 1000 km to 6000 km in altitude (corresponding to $\sim (0.2$--$1)\,R_E$ above Earth’s surface) and spanning L-shell values of $L \sim 1.2$--2. This belt predominantly consists of energetic protons with energies exceeding 100~MeV \citep{1}, trapped by Earth's strong geomagnetic field \citep{B3}, and is therefore commonly referred to as the proton belt. The energetic proton population is sufficiently low in density to be considered effectively collisionless. Under the influence of the Lorentz force, charged particles in the geomagnetic field execute three characteristic types of motion: gyromotion about the magnetic field lines, bounce motion between magnetic mirror points near the polar regions, and azimuthal drift motion around the Earth. The particle density in the lower magnetosphere is often modeled using an exponential profile for oxygen ions combined with a power-law distribution for hydrogen ions \citep{11}, and is expressed as:

\begin{equation}\label{obs_chargedensity}
 n(r)= {n_O} {e^{-(r-R_I)/h}} +{n_H}r^{-1}
\end{equation}

where $r$ denotes the geocentric radial distance, $n_O = 10^{5}\,\mathrm{cm^{-3}}$ and $n_H = 10^{3}\,\mathrm{cm^{-3}}$ are the reference densities of oxygen and hydrogen ions, respectively, $h = 400\,\mathrm{km}$ is the scale height, and $R_I = 1.0314\,R_E$, with $R_E$ being Earth's radius. The radial density profile described by Eq.~\eqref{obs_chargedensity} is presented in Fig.~\ref{fig:observed_density}.

Based on the above description, we now develop a kinetic model of high-energy trapped protons in the lower magnetosphere. We obtain a steady-state kinetic distribution of energetic particles that reproduces the observed radial density profile of the inner Van Allen radiation belt (Fig.~\ref{fig:observed_density}). We first construct an initial $f(E,\mu)$ distribution function analytically in Section~\ref{section:analytical}, and subsequently use it to perform numerical simulations of particle dynamics along geomagnetic field lines, considering different combinations of perpendicular and parallel energy with respect to the mean magnetic field direction, as described in Section~\ref{section:numerical}. In the following subsections, we present the distribution functions obtained from these simulations that successfully reproduce the observed density profile.

\subsection{Analytical model} \label{section:analytical}

The particle number density is obtained by integrating the phase space distribution function, $f(\mathbf{v})$, over the velocity space. In spherical polar coordinates ($v,\theta,\phi$), this is expressed as:

\begin{equation}\label{am1}
n=\int \int \int f(v,\theta,\phi)v^2 \sin\theta dv d\theta d\phi
\end{equation}    

where $v$ is the magnitude of the particle velocity, $\theta$ is the polar angle between the particle's velocity vector and the direction of the magnetic field, and $\phi$ is the azimuthal angle describing the orientation of the particle's velocity in the plane perpendicular to the magnetic field.

The particle energy $E$,  and the magnetic moment $\mu$,  are defined as: 

\begin{equation}\label{am2}
 E = \frac{1}{2}mv^2 = \frac{1}{2}m{v_\parallel}^2 + \frac{1}{2}m{v_\perp}^2,  \mu=\frac {mv_\perp ^ 2}{2B} , \quad
\end{equation}  

Here, $B$ is the magnetic field strength, 
$v_\parallel$is the component of the particle velocity parallel to the magnetic field, and $v_\perp$ is the component perpendicular to the magnetic field. The magnetic moment $\mu$ represents the particle’s gyration around the field line and is an adiabatic invariant in slowly varying fields. The total energy of a particle can be rewritten as:

\begin{equation}\label{am3}
E= \frac{1}{2}m{v_\parallel}^2 + \mu B
\end{equation}

with:
\begin{equation}\label{am4}
v_\perp = v \sin\theta,  v_\parallel=v \cos\theta, \quad
\end{equation}

The infinitesimal variations of the quantities $E$, $\mu$ and $v_\perp$ are:
\begin{equation}\label{am5}
dE=mvdv ; d\mu=\frac {mv_\perp dv_\perp}{B}; dv_\perp=v \cos\theta d\theta; 
\end{equation}

We express Eq.~\ref{am1} in terms of $E$ and  $\mu$ by substituting Eqs.~\ref{am2}, \ref{am3}, \ref{am4} and \ref{am5} into it. The resulting form of Eq.~\ref{am1} is:

\begin{equation}\label{am6}
 n=\frac{1}{\sqrt{2}m^{3/2}}\int \int \int \frac{ f(E,\mu) B}{\sqrt{E-\mu B}}dE d\mu d\phi
\end{equation}

Now, we use the expression in Eq.~\eqref{am6} to calculate the particle density distribution along a magnetic field line at $L = 1.5$, with the magnetic field strength expressed as $B(s)$, where $s$ denotes the position along the field line. The integration over the azimuthal angle $\phi$ is performed from $0$ to $2\pi$ to account for the full circular motion around the magnetic field, as the integrand is independent of $\phi$. The magnetic moment $\mu$ is integrated from $0$ to $E / B(s)$, since the perpendicular energy cannot exceed the total particle energy, ensuring that the square root in the denominator remains real. The particle energy $E$ is integrated from $0$ to $\infty$ to include all possible particle energies, thereby covering the entire velocity space.
With these limits, the particle number density along the field line can be expressed as:

\begin{equation}\label{eq:mfc}
 n(s)=\frac{1 }{\sqrt{2}m^{3/2}} \int^\infty_0  \int^\frac{E}{B}_0 \int^{2\pi}_0\frac{f(E,\mu) B(s)}{\sqrt{E-\mu B(s)}}dE d\mu d\phi
\end{equation} 

The final expression after integrating over $\phi$ is:

\begin{equation}\label{eq:mfc}
 n(s)=\frac{2\pi B(s)}{\sqrt{2}m^{3/2}}\int^\infty_0 \int^\frac{E}{B}_0 \frac{f(E,\mu) }{\sqrt{E-\mu B(s)}}dE d\mu
\end{equation}

To calculate $n(s)$ from $f(E,\mu)$ in Eq.~\ref{eq:mfc} , we assume that the phase-space distribution function is separable in energy and magnetic moment, i.e.,

\begin{equation}\label{eq:mfs}
 f(E,\mu)=f(E)f(\mu)
\end{equation}   

\textcolor{black}{This assumption in Eq.~\ref{eq:mfs} is adopted as a simplifying model for the initial trapped particle population and allows us to independently control the energy spectrum and pitch-angle anisotropy. In a collisionless magnetospheric plasma, energy $E$ and magnetic moment $\mu$ are conserved adiabatic invariants (in the absence of wave–particle interactions). It is therefore reasonable to prescribe their initial distributions independently.}
 
The energetic protons in the inner Van Allen belt are effectively collisionless. The chosen forms correspond to Gamma-type distributions (exponential with polynomial prefactor)  and are introduced phenomenologically to represent a steady trapped particle population. These distributions retain exponential (Boltzmann-type) decay in both energy and magnetic moment, while allowing flexible shaping of their dependence through polynomial prefactors, thereby introducing modified weighting in both variables and enabling deviations from a collisional Maxwellian. We find that the following distribution function f($E,\mu$) chosen in Eq.~\ref{iemu} approximately resembles the observed density in the inner radiation belt.

\begin{equation}\label{iemu}
  f(E)=E e ^  \frac{-E}{KT} ,\qquad f(\mu) = \mu e ^{-\mu}
\end{equation}

The density profile obtained from this predicted distribution function is presented later in Sec 2.2.4  as a comparison with observed density and density profile obtained from numerical simulations.

\subsection{Numerical model}\label{section:numerical}

\subsubsection{Dipole model of the Geomagnetic Field at $L = 1.5$}\label{section:dipole}

The inner radiation belt is estimated to be located between L-shell 1.2 to 2. \textcolor{black}{The L-shell parameter, $L$, characterizes a particular magnetic field line in the Earth's dipole field. It is defined as the radial distance (in Earth radii) from the center of the Earth to the point where that field line crosses the geomagnetic equatorial plane.} We choose \textcolor{black}{an} L-shell value of 1.5 for modeling the inner radiation belt. \textcolor{black}{
The IITMSAT orbit is designed to cover the L-shell range $L \approx 1.15$--$1.75$ at altitudes of 600--800~km, corresponding to magnetic latitudes of approximately $17^\circ$--$40^\circ$, which defines the region where precipitation from the inner radiation belt can be effectively detected. The value $L = 1.5$ is therefore chosen as an intermediate shell near the center of this observable range and within the inner Van Allen radiation belt ($L \approx 1.2$--$2.0$). This choice avoids very low L-shells (e.g., $L \approx 1.2$), where the equatorial magnetic field is stronger, resulting in a larger loss cone and weaker trapping, thereby reducing the steady trapped population in our numerical simulations and limiting trapped-particle statistics. Conversely, larger L values ($L \ge 2$) approach the transition toward the slot region (the depleted energetic-particle region between the inner and outer radiation belts), where particle dynamics differ from those of the inner-belt proton population. In addition, precipitation from higher L-shells occurs at larger magnetic latitudes, beyond the mission coverage limit of $40^\circ$, reducing the likelihood of detection by IITMSAT. Since $L = 1.5$ corresponds to an intermediate latitude ($\sim 30^\circ$) within the mission coverage band, the satellite repeatedly intersects this shell during successive orbits, improving statistical sampling and detection probability. Therefore, $L = 1.5$ provides a physically suitable location for modeling wave--particle interactions and predicting detectable energetic particle precipitation relevant to the IITMSAT mission.}

We also choose a simplified dipole model of the Earth's magnetic field which holds good for lower L shells and is a first order approximation of the rather complex true Earth's magnetic field. The dipole model is described in polar $\textbf{B}(r,\textcolor{black}{\theta_{GM}})$ coordinates, where $r$ is the geocentric radius, \textcolor{black}{$\theta_{GM}$} is the geomagnetic latitude  measured northward from the equator, and $s$ is the arc length along the L-shell, related as follows:   

\begin{equation}\label{Lshell_eqn}
\begin{aligned}
r = Lcos^2\textcolor{black}{\theta_{GM}} , \quad ds^2=dr^2 +(rd\textcolor{black}{\theta_{GM}})^2 , \\
\quad d\theta_{GM}=\frac{ds}{L\sqrt{sin^2 2\textcolor{black}{\theta_{GM}}+cos^4\textcolor{black}{\theta_{GM}}}}
\end{aligned}
\end{equation}  

\begin{figure}[h]
\centering
  \includegraphics[width=0.8\linewidth]{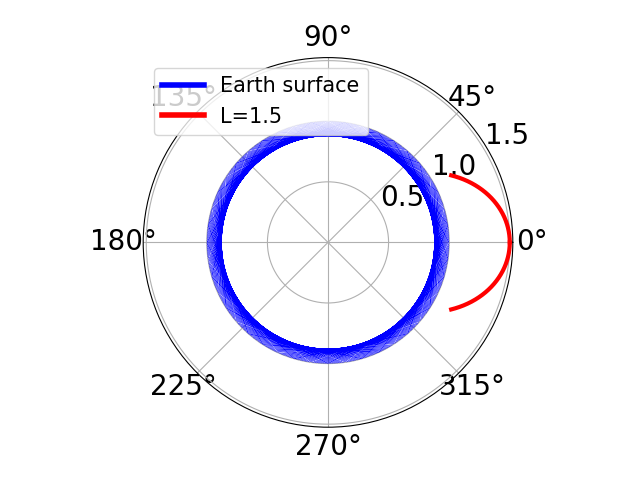}
  \vspace{1em}
 \caption{A polar plot of the lower magnetosphere L shell 1.5. The blue thick line represents the \textcolor{black}{Earth's surface} and the red line represents the L-shell 1.5. In this representation, the L-shell (red line) is truncated at an altitude of 1000 km above the Earth’s surface, which is taken as the top boundary of the ionosphere. Particles crossing this boundary are considered to be detrapped or precipitated from the radiation belt.}
  \label{fig:dipole_mf_earth}
\end{figure}

A polar representation of the L-shell 1.5 is presented in Fig.~\ref{fig:dipole_mf_earth}, constructed using Eq.~\ref{Lshell_eqn}. 

In this plot, the topside of the ionosphere is taken to be at an altitude of 1000 km above the Earth’s surface and is treated as the effective mirror (loss-cone) boundary for particles in the inner radiation belt. Particles crossing this boundary are considered to be detrapped or lost via precipitation. This boundary defines the extent of the $s$-coordinate system, where $s$ denotes the arc length along the $L$-shell, assumed to be symmetric about the geomagnetic equator. The dipole magnetic field strength in polar coordinates is given as:

\begin{equation}\label{mf_strength}
   B(r)=\textcolor{black}{B_0}(\frac{R_E}{r})^3 \sqrt{(1+3\sin^2\textcolor{black}{\theta_{GM}})}
\end{equation}  

\textcolor{black}{where $B_0$ is the magnetic field at the magnetic equator on the Earth's surface, with a typical value of 0.312$\mu$T.}

After developing the geomagnetic dipole field model at the $L = 1.5$ shell, we construct the initial particle distribution functions used to obtain a steady-state trapped population in the inner radiation belt. The initial distribution consists of the spatial particle density $n(s)$ along the $L = 1.5$ shell, and the energy–magnetic moment distribution $f(E,\mu)$.

\subsubsection{Initial Particle Distribution}\label{section:spatial}

\textcolor{black}{Energetic} particles are initially distributed uniformly along the one-dimensional (1D) field-aligned coordinate '$s$'. \textcolor{black}{This choice of the uniform distribution represents a bounce-phase–mixed ensemble that undergoes rapid bounce-phase mixing of energetic particles on the $L=1.5$ shell. Over a few bounce periods, particles with different energies and magnetic moments sample all accessible positions between their mirror points, providing an unbiased sampling of the accessible phase space and a physically justified representation of the trapped population along the field line.} This one-dimensional spatial (1D) distribution is transformed into a three-dimensional (3D) spatial distribution $f(\mathbf{r}) \equiv f(x,y,z)$ in Cartesian coordinates as:

\begin{equation}\label{eq:ic_posn}
x=r\cos\textcolor{black}{\theta_{GM}}, y=0,  z=r \sin \textcolor{black}{\theta_{GM}}
\end{equation} 

This spatial distribution $f(\mathbf{r}) \equiv f(x,y,z)$ is used as the initial condition in the simulations described later to evolve particle trajectories along the inner radiation belt using the Lorentz force equation (see Eqs.~\ref{LF}).

Next an initial $f(E, \mu)$ distribution is chosen which is similar in form to  Eq.~\ref{iemu} as predicted in the analytical model. We consider the simplest assumption that $f(E)$ and $f(\mu)$ are independent of each other, which allows us to independently control the energy spectrum and pitch-angle anisotropy in our numerical setup. The energy distribution $f(E)$ is chosen as in Eq.~\ref{iemu} with KT = 50 MeV. 

In our previous study, \citep{32} the magnetic moment distribution, $f(\mu)$, was derived directly from the same energy variable $E$ used in the distribution function $f(E)$, i.e.,

\begin{equation}\label{eq:icp}
f(\mu)= \frac{f(E)}{\alpha B_{max}} 
\end{equation}

 \textcolor{black}{Here, $B_{max}$ is the maximum value of the  static geomagnetic field, representing the mirror-point magnetic field, and $\alpha$ is a dimensionless parameter that controls the ratio between  perpendicular ($E_\perp$) and parallel energy ($E_\parallel$) in the initial particle distribution. While $\alpha$ is not the pitch angle itself, it effectively determines the typical pitch-angle range of particles by setting how much of their energy is perpendicular versus parallel to the magnetic field. Varying $\alpha$ changes the initial anisotropy in the $f(E, \mu)$ distribution and consequently leads to different steady-state density profiles in the inner radiation belt.} 

 Since, the magnetic moment distribution was derived directly from the same energy variable $E$, this prescription established a one-to-one correspondence between  $E$ and $\mu$.  Thus, although $E$ and $\mu$ are independent adiabatic invariants in a collisionless magnetized medium, the adopted formulation imposed a deterministic relation between them. As a result, the two-dimensional phase space was effectively reduced to a one-dimensional manifold, and particles were arranged linearly in the $E–\mu$ space, producing a diagonal structure in the phase-space distribution\citep{32}. The value of $\alpha$ was varied to generate different initial $f(E,\mu)$ distributions, which in turn led to different steady-state density profiles and reproduced the observed density structure of the inner VA belt.

In the present study, we adopt a modified prescription for the initial $f(E,\mu)$ distribution in order to remove the imposed constraint of one-to-one correspondence between energy and magnetic moment and allow particles to occupy a finite region in the $E–\mu$ space. The magnetic moment distribution, $f(\mu)$, is now prescribed as:

\begin{equation}\label{eq:ic_Enew}
f(\mu)= \frac{f(E')}{\alpha B_{max}} 
\end{equation}
    
Here, $f(E')$ has exactly the same functional form and parameters as the energy distribution $f(E)$, but is evaluated using an independent energy variable $E'$. The analytical form is therefore unchanged, ensuring consistency with the adopted near-Maxwellian distribution. Independent sampling of $E$ and  $E'$  from the same near-Maxwellian distribution removes the one-to-one constraint between $E$ and  $\mu$, thereby permitting independent variation of energy and magnetic moment and resulting in a finite spread of particles in the $ E–\mu$ space. 

\begin{figure}[h]
\centering
\includegraphics[width=0.8\linewidth]{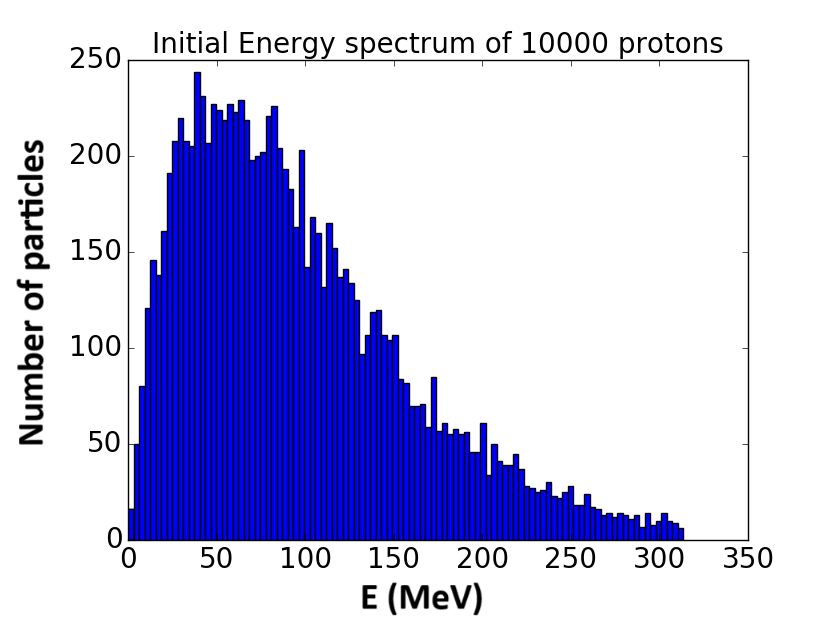}  
  \vspace{1em}
 \caption{ Initial energy spectrum of 10000 protons distributed along the L shell 1.5, with KT = 50 MeV. The $x$-axis shows particle energy in MeV, while the $y$-axis represents the continuous distribution function, indicating the number of particles corresponding to each energy value.}
  \label{fig:inie}
\end{figure}

The initial distributions $f(E)$, $f(\mu)$, and $f(E,\mu)$, generated numerically and employed in our simulations to reproduce the inner radiation belt density at $L = 1.5$, are presented next. These distributions correspond to a chosen value of $\alpha = 10$, which provides the best fit to the observed density. The initial $f(E)$ is presented in {Fig.~\ref{fig:inie}}. The distribution resembles the form of $f(E)$ given in Eq.~\ref{iemu}, corresponding to a gamma-like function with $kT = 50,\text{MeV}$, constructed using $10^4$ particles.

\begin{figure}[h]
\centering
\includegraphics[width=0.8\linewidth]{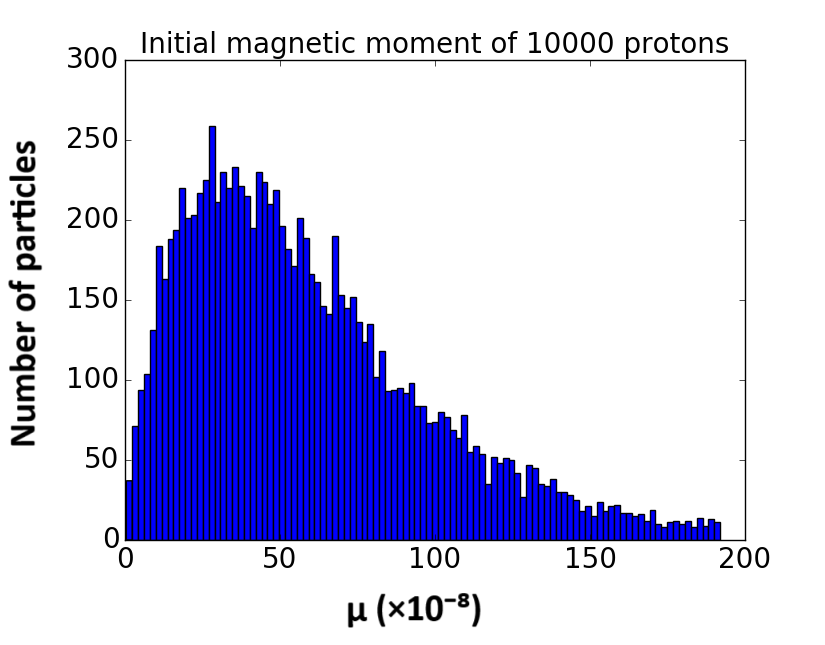}  
  \vspace{1em}
 \caption{Initial magnetic moment distribution of 100000 particles along L shell 1.5. The shape of this distribution  is similar to that of the energy distribution function.}
  \label{fig:inimu}
\end{figure}

The initial $f(\mu)$ is shown in {Fig.~\ref{fig:inimu}}. The distribution is obtained using Eq.~\ref{eq:ic_Enew} and closely resembles the form of $f(\mu)$ given in Eq.~\ref{iemu}, corresponding to a gamma-like function constructed using $10^4$ particles.

\begin{figure}[h]
\centering
\includegraphics[width=0.8\linewidth]{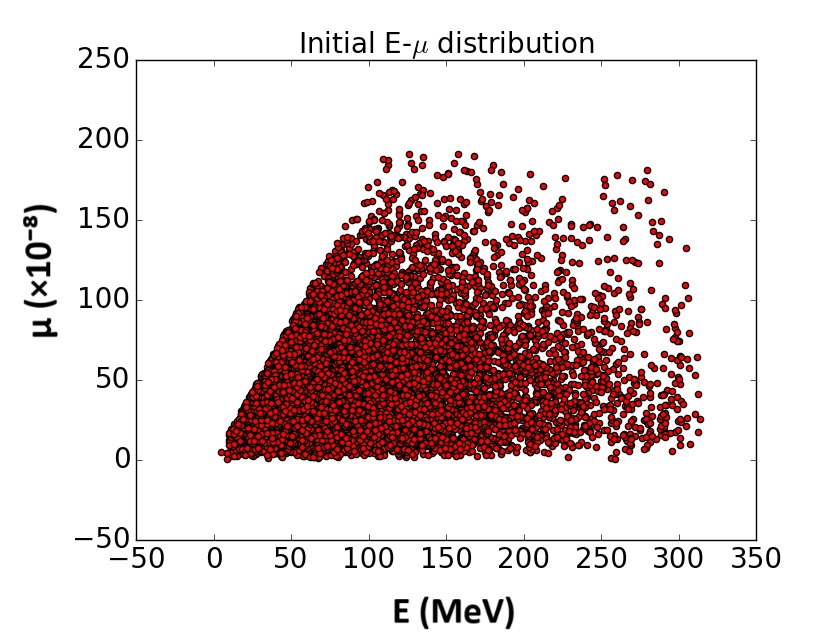}  
  \vspace{1em}
 \caption{Initial f(E,$\mu$) from Eq.~\ref{iemu} and Eq.~\ref{eq:ic_Enew} which is a product of two independent gamma functions. This plot is for KT = 50 MeV and $\alpha = 10$. The $x$-axis represents the energy of particles in MeV and the $y$-axis represents values of $\mu$.}
  \label{fig:iemu}
\end{figure}

The initial $f(E, \mu )$ for $\alpha=10$ is presented in {Fig.~\ref{fig:iemu}}  which results in the observed particle density in the inner radiation belt. It is seen that the current $f(E,\mu)$ distribution generated here occupies a finite two-dimensional E-$\mu$ space compared to our earlier choice of distribution function.

To initialize particle velocities for the simulations, the energy $E$ and magnetic moment $\mu$ are converted into parallel and perpendicular velocity components. The perpendicular velocity is obtained from the magnetic moment, while the total velocity is determined from the particle energy. The perpendicular velocity is then decomposed into Cartesian components using a uniformly distributed gyro-phase angle $\phi$, yielding the three-dimensional velocity components $v_x$, $v_y$, and $v_z$ as:

\begin{equation}\label{eq:ic_vel}
\begin{aligned}
   v_\perp=\sqrt{\frac{2 \mu B}{m}}  , \quad
   v_x=v_\perp \cos \phi , v_y=v_\perp \sin \phi , \\
   v_z= \sqrt{v^2-{v_x}^2-{v_y}^2}
\end{aligned}
\end{equation}  

This transformation results in the corresponding three-dimensional velocity-space distribution $f(\mathbf{v}) \equiv f(v_x,v_y,v_z)$, which is used as the initial condition in the simulations described later to evolve particle trajectories along the inner radiation belt using the Lorentz force equation (see Eqs.~\ref{LF}).

\subsubsection{Particle Dynamics and Numerical Solver}

Next, we carry out particle simulations of energetic protons, initialized using the constructed distribution functions (Eqs.~\ref{iemu}, \ref{eq:ic_posn}, \ref{eq:ic_Enew}, and \ref{eq:ic_vel}), to obtain a steady-state trapped particle population in the magnetic mirror configuration of the inner radiation belt that reproduces the observed density profile of the inner radiation belt. The initial particle distribution is specified in three-dimensional Cartesian coordinates. The particles are assigned a uniform spatial distribution, which is transformed into a three-dimensional spatial distribution $f(\mathbf{r}) \equiv f(x,y,z)$ using Eq.~\ref{eq:ic_posn}. The initial velocity distribution is prescribed through a distribution function $f(E,\mu)$ (see Fig.~\ref{fig:iemu}), which is subsequently transformed into a three-dimensional velocity-space distribution $f(\mathbf{v}) \equiv f(v_x,v_y,v_z)$ using Eq.~\ref{eq:ic_vel}. The particle trajectories are then evolved by solving the Lorentz force equation:

\begin{equation}\label{LF}
\begin{aligned}
 m\frac{d \textbf{v}}{dt} = q \textbf {v} \times  {{\textbf{B}_{static}}} \\
   \end{aligned}
\end{equation}

where q, m and \textbf{v} represent the charge, mass and velocity of the particles respectively.  ${\textbf{B}_{static}}$ is the 
steady state dipole geomagnetic field of the earth without any MHD disturbance in 3D Cartesian coordinates and is given by:

\begin{equation}
\begin{aligned}
   B_{x}= \frac{-3xz B_0 (R_E)^3}{(x^2 +y^2 +z^2)^\frac{5}{2}} ;
   B_{y}= \frac{-3yz B_0 (R_E)^3}{(x^2 +y^2 +z^2)^\frac{5}{2}} ;\\
   B_{z}= \frac{-(2z^2-x^2-y^2) B_0 (R_E)^3}{(x^2 +y^2 +z^2)^\frac{5}{2}}
\end{aligned}
\end{equation}

\textcolor{black}{where $B_0$ is the magnetic field at the magnetic equator on the Earth's surface, with a typical value of 0.312$\mu$T.}

We use 10000 protons to carry out our inner radiation belt modeling simulations. The Lorentz force equation (Eq.~\ref{LF}) is solved using the fourth-order Runge--Kutta (RK4) method for particle trajectory integration. The RK4 scheme has a local truncation error of $\mathcal{O}(h^5)$ and a global accuracy of $\mathcal{O}(h^4)$, and is chosen for its high accuracy and numerical stability. The particle solver is parallelized using OpenMP to efficiently evolve a large number of particle trajectories.  For the numerical simulations, the discretized time step used for particle advancement in solving the ODEs is chosen such that it satisfies the Nyquist criterion and accurately resolves the particle trajectories. \textcolor{black}{The gyroperiod, $T_{\text{gyro}}$, of the particles is given by
\begin{equation}
T_{\text{gyro}} = \frac{2\pi m_p}{qB},
\end{equation}
where $m_p$ and $q$ are the proton mass and charge, respectively, and $B$ is the local magnetic field strength. Along the $L$-shell, the magnetic field varies, and consequently, the gyroperiod also varies, attaining its smallest value near the magnetic poles where the field is strongest. The simulation must therefore resolve this minimum gyroperiod to accurately capture particle dynamics. In the inner radiation belt (near $L \sim 1.5$), the magnetic field strength near the poles (or mirror points) is typically $B \sim 25\text{--}26~\mu\text{T}$. Substituting these values, we obtain
\begin{align}
T_{\text{gyro}} &= \frac{2 \pi (1.673\times10^{-27}\ \text{kg})}
{(1.602\times10^{-19}\ \text{C}) (26\times10^{-6}\ \text{T})} \\
&\approx 2.52 \times 10^{-3}\ \text{s} = 2500\ \mu\text{s}.
\end{align} }

\textcolor{black}{This corresponds to the smallest proton gyroperiod, as it occurs at the maximum magnetic field encountered along the field line. For typical inner radiation belt conditions, this may be rounded to $\sim 2000~\mu\text{s}$ as an order-of-magnitude estimate. Thus, the smallest gyroperiod is dictated by the strongest magnetic field in the inner belt, which sets the fastest cyclotron motion for trapped protons. Based on this, we adopt a discretized time step of $10~\mu\text{s}$, which sufficiently resolves the gyro-motion and satisfies the Nyquist criterion in our simulations.}
 
\begin{figure}
\centering

\includegraphics[width=0.8\linewidth]{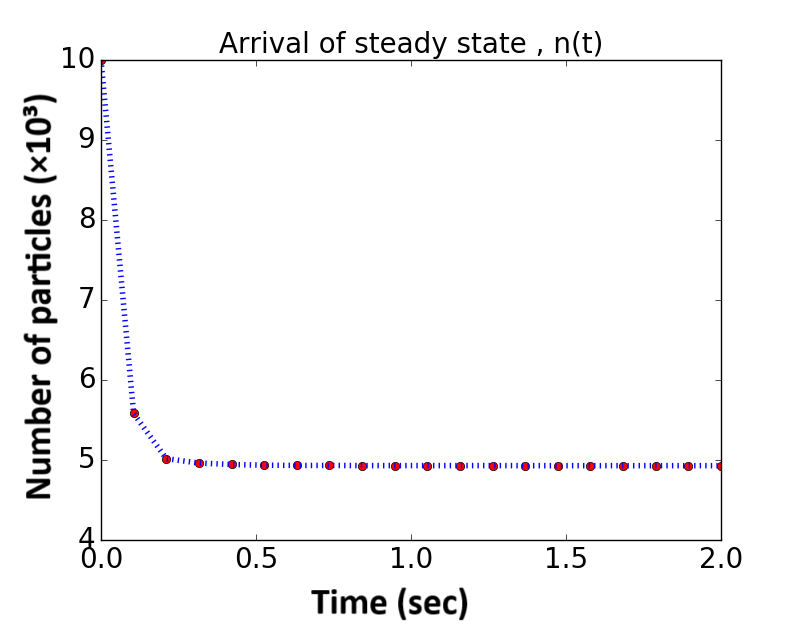}  
  \vspace{1em}
 \caption{Temporal evolution of the number of trapped protons in a magnetic mirror configuration, starting from an initial ensemble of 10,000 particles with a prescribed (E, $\mu$) distribution. As particles bounce between mirror points along the geomagnetic field lines, those with smaller magnetic moment $\mu$ (corresponding to larger parallel and smaller perpendicular velocities) enter the loss cone and are progressively lost to the ionosphere. The system rapidly evolves toward a steady state, with approximately 4,820 particles remaining trapped. Given a characteristic proton bounce period of ~0.2 s, the steady state is established within only a few bounce cycles.}
  \label{nt}
\end{figure}

During the simulations, particles undergo gyromotion about the magnetic field lines, execute bounce motion along the geomagnetic $L = 1.5$ field line, reflecting between magnetic mirror points formed near the polar regions, and undergo azimuthal drift around the Earth.  The dipolar geomagnetic field strength increases toward the polar regions along the $L = 1.5$ field line. As charged particles move along the field, conservation of the magnetic moment $\mu = mv_\perp^2/2B$ requires an increase in perpendicular velocity as $B$ increases. This leads to a corresponding decrease in the parallel velocity, eventually causing it to vanish and reverse direction, resulting in reflection at magnetic mirror points near the poles. However, particles  with very small initial pitch angle are lost from the mirror and the rest of the particles attain a steady state in our simulation (see Fig.~\ref{nt}). 

\textcolor{black}{In the present model, the "mirror-point" logic is used to determine particle loss from trapped orbits. In our simulations, particles that cross the lower magnetospheric boundary (i.e., the magnetosphere–ionosphere interface) are treated as lost and are no longer dynamically evolved. Specifically, when a particle reaches the prescribed mirror altitude ($\sim$1000 km, corresponding to the topside ionosphere), its motion is frozen and the particle is subsequently removed from the simulation. In the numerical implementation, this is modeled through a stochastic (Monte Carlo-type) absorbing boundary condition. Specifically, when a particle reaches a position satisfying $|z| > z_{\max}$, where $z_{\max}$ corresponds to the chosen precipitation altitude, it is assigned a probability of removal given by
\begin{equation}
P_{\mathrm{loss}} = 1 - \frac{1}{1 + \exp\left[\kappa \left(|z| - z_{\max}\right)\right]},
\end{equation}
where $\kappa$ is a parameter controlling the sharpness of the transition. A random number $r \in [0,1]$ is then drawn, and the particle is considered lost (precipitated) if $r < P_{\mathrm{loss}}$. Numerically, this is implemented by freezing the particle’s trajectory through a flag that suppresses further updates in the integration scheme. This approach provides a smooth, probabilistic representation of particle loss near the boundary, avoiding artificial discontinuities associated with a hard cutoff.}
 
 \textcolor{black}{Atmospheric backscatter and quasi-trapped particle populations are not included in the current model. This simplification is adopted to focus on first-order precipitation driven by wave–particle interactions and to avoid introducing additional uncertainties associated with atmospheric scattering processes. The adopted approach therefore treats particles entering the loss cone as precipitating and permanently lost from the trapped population. While atmospheric backscatter can, in principle, produce quasi-trapped particles, incorporating such effects would require a more detailed atmosphere–ionosphere interaction model and is not addressed in the present study.} 

The simulations are performed for a duration of 2 s for protons. Figure~\ref{nt} shows the temporal evolution of the number of trapped particles in a magnetic mirror configuration, starting from an initial ensemble of $10^4$ particles with a prescribed $(E,\mu)$ distribution. As particles bounce between mirror points along geomagnetic field lines, those with smaller magnetic moment $\mu$ (corresponding to larger parallel and smaller perpendicular velocities) enter the loss cone and are progressively lost to the ionosphere. The system rapidly evolves toward a steady state, with particles achieving saturation within $\sim 0.2$–0.3 s, which is comparable to the characteristic proton bounce period ($\approx 0.2$ s). This indicates that particles with small pitch angles are lost within one or two bounce cycles, after which the remaining population remains stably trapped. At steady state, approximately 4,820 particles remain confined out of the initial $10^4$, indicating that the steady-state configuration is established within only a few bounce cycles.

\subsubsection{Steady State Particle Distribution}\label{section:spatial}

\begin{figure}[h]
\centering
\includegraphics[width=0.8\linewidth]{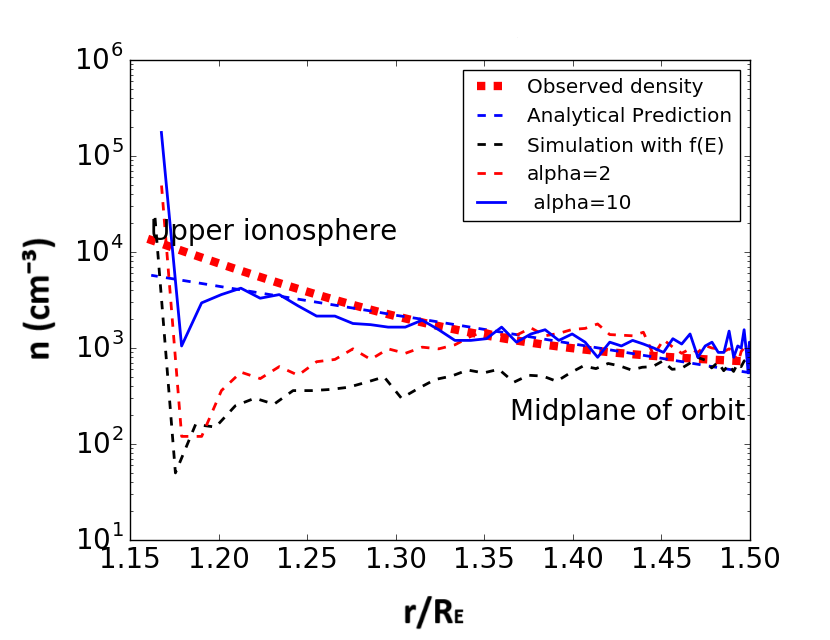}  
  \vspace{1em}
 \caption{Density profile obtained numerically for different values of $\alpha$ and compared with observations. Figure also shows comparison with analytical calculations. The analytical calculations takes all particles into account and does not distinguish trapped particles from the lost ones whereas the numerical simulations takes only trapped particles into account for modelling observed density and hence more accurate. Simulations of particles designed with just $f(E)$ as initial conditions does not give observed density profile. An $\alpha$ value of 10 gives the best results for obtained density.}
  \label{fig:comp_density}
\end{figure}

 Fig.~\ref{fig:comp_density} compares the observed, analytically predicted, and numerically obtained densities from simulations performed for different values of the parameter $\alpha$. The best agreement with the observed density is obtained for $\alpha = 10$. In our earlier studies \citep{32} a best fit for $\alpha=20$ was obtained. The choice of the modified $f(E,\mu)$ distribution function in the present simulations leads to a reduced anisotropy in the particle energy distribution. For sufficiently large $\alpha$, the system develops a strong temperature (or pressure) anisotropy, with the parallel temperature exceeding the perpendicular temperature. Such anisotropy provides free energy for kinetic instabilities such as the firehose instability, and very large $\alpha$ values may therefore drive the system toward instability, potentially leading to magnetic field line distortion or buckling.

\textcolor{black}{We note that in the actual inner Van Allen radiation belt, particle anisotropy can vary under geomagnetic-storm and solar-wind driving, and sufficiently strong anisotropy may in principle excite such instabilities, modifying the wave environment. However, these self-consistent anisotropy-driven plasma responses are not included in the present model. In our formulation, the parameter $\alpha$ is introduced phenomenologically in the initial $f(E,\mu)$ distribution to control the relative perpendicular and parallel energy content of the trapped proton population. Its role in the present study is to generate different initial anisotropies and determine which choice reproduces the observed steady-state density profile of the inner belt most accurately. The value $\alpha = 10$ provides better agreement with the observed density profile while maintaining a moderate anisotropy and reducing the possibility of instability growth, such as the firehose instability, within the limits of the present model.}

\textcolor{black}{The present work does not attempt to model storm-time evolution of anisotropy or the self-consistent excitation of instabilities and their feedback on the wave spectrum and precipitation rates. Instead, the focus is on constructing a representative steady trapped population at $L=1.5$ and studying its interaction with prescribed Alfv'en-wave disturbances. The choice $\alpha = 10$ should therefore be interpreted within this restricted modeling framework as the value that best reproduces the steady-state density profile while maintaining a moderate anisotropy consistent with the assumptions of the present kinetic model.}

\textcolor{black}{We note that the apparent enhancement in particle density within the radial range $1.15 < r/R_E < 1.20$ in Fig.~\ref{fig:comp_density}, is a numerical artifact rather than a physical transition region. In our simulations, particles that cross the lower magnetospheric boundary (i.e., the magnetosphere–ionosphere interface) are treated as lost and are no longer dynamically evolved. However, these particles were not removed from the density accumulation used to construct the radial profile. As a result, they artificially accumulate near the boundary, producing the observed spike in density.}

\begin{figure}[h]
\centering
\includegraphics[width=0.8\linewidth]{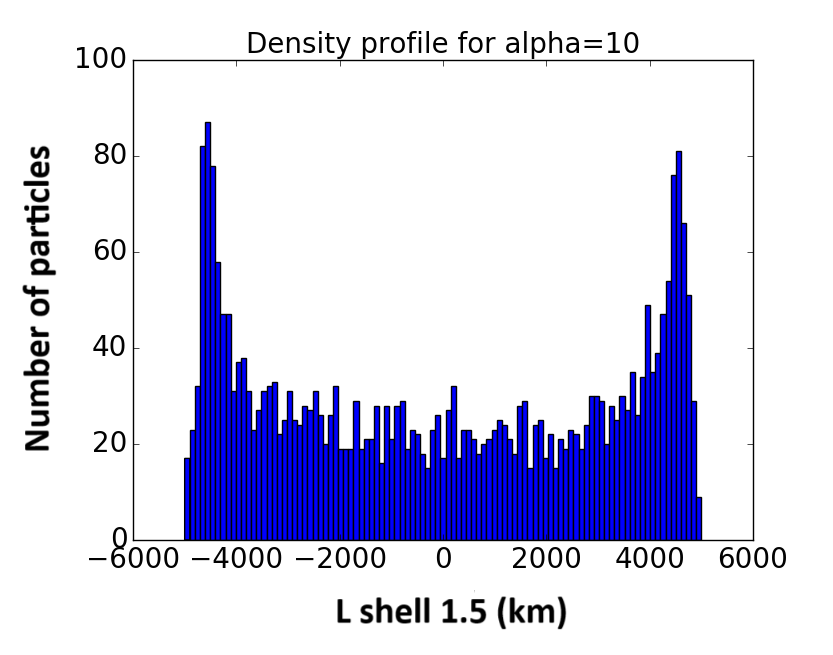}  
  \vspace{1em}
 \caption{Density distribution for $\alpha = 10$ along L = 1.5. This profile best reproduces the observed density, with more particles concentrated near the edges of the $s$ coordinate compared to the $\alpha = 2$ case (see Fig.~\ref{alpha2}).}
  \label{fig:density_alpha10}
\end{figure}

\begin{figure}
\centering
\includegraphics[width=0.8\linewidth]{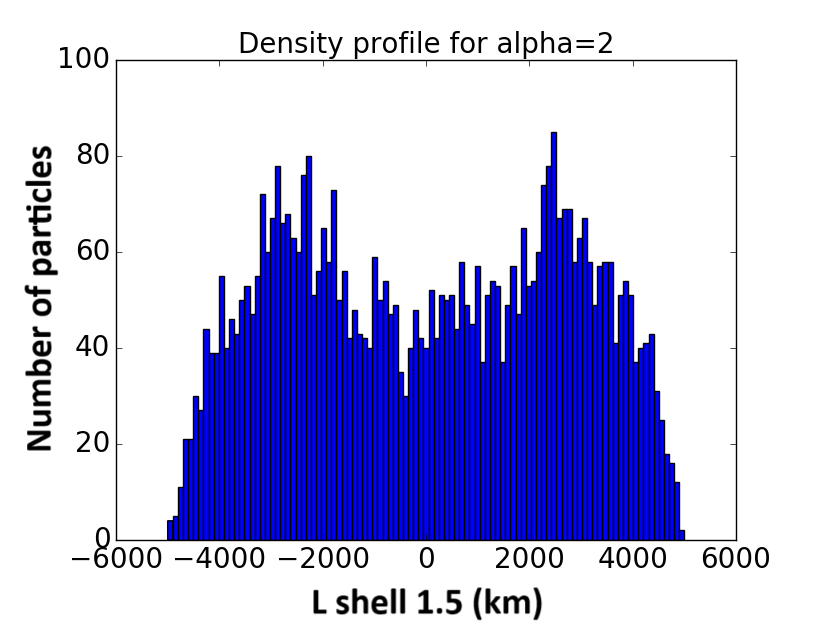}  
  \vspace{1em}
 \caption{Density distribution for $\alpha = 2$ along L = 1.5. This profile is less similar to the observed density because particles have lower parallel energy, resulting in shorter bounce amplitudes and a peak near ~3000 km. Increasing $\alpha$ shifts the distribution outward, producing larger bounce amplitudes and a peak near ~5000 km, which better matches observations (see Figs.\ref{fig:comp_density} and  ~\ref{fig:density_alpha10}).}
  \label{alpha2}
\end{figure}

At steady state, the density profile $n(s)$ for $\alpha = 10$ is shown in Fig.~\ref{fig:density_alpha10}. For comparison, the corresponding profile for $\alpha = 2$ is presented in Fig.~\ref{alpha2}, where a more pronounced deviation from the observed density is evident. Overall, $\alpha = 10$ provides the best agreement with the observed density profile. 

For the lower value, $\alpha = 2$, particles possess comparatively smaller parallel energy, resulting in shorter bounce amplitudes along the magnetic field lines and stronger confinement toward the central region. Consequently, the particle population is enhanced around $\sim 3000$~km, as observed in the simulations. In contrast, increasing $\alpha$ enhances the parallel energy component, leading to larger bounce amplitudes and allowing particles to access regions farther along the field line. As a result, the spatial distribution shifts outward and peaks at higher altitudes ($\sim 5000$~km), closer to the edges of the $s$-coordinate along the $L = 1.5$ shell.

\begin{figure}[h]
\centering
\includegraphics[width=0.8\linewidth]{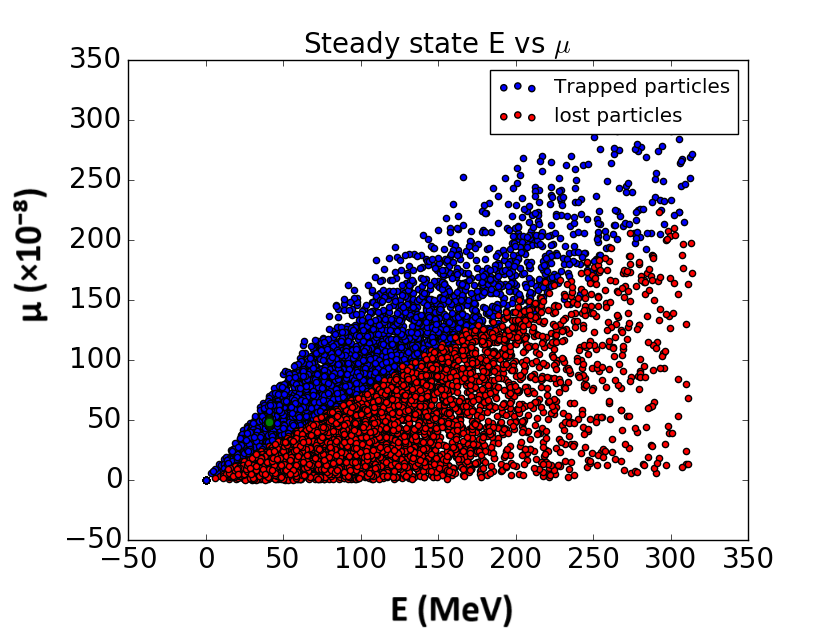}  
  \vspace{1em}
 \caption{Steady state $f(E,\mu) $ from numerical simulations for $\alpha=10$. The particles marked by red are in a loss cone whereas the blue ones are trapped. There is an overlapping region where the gyrophase decides if particles are trapped or lost. The $x$-axis represents particle energies in MeV and the $y$-axis represents values of $\mu$.} 
  \label{fig:femu}
\end{figure}

The $E-\mu$ distribution at steady state is presented in Fig.~\ref{fig:femu} for $\alpha$=10. The lost particles (in the loss cone) clearly separates out from the trapped particles in orbit in Fig.~\ref{fig:femu} when $\mu$ is observed at their bounce points.

\begin{figure}[h]
\centering
\includegraphics[width=0.8\linewidth]{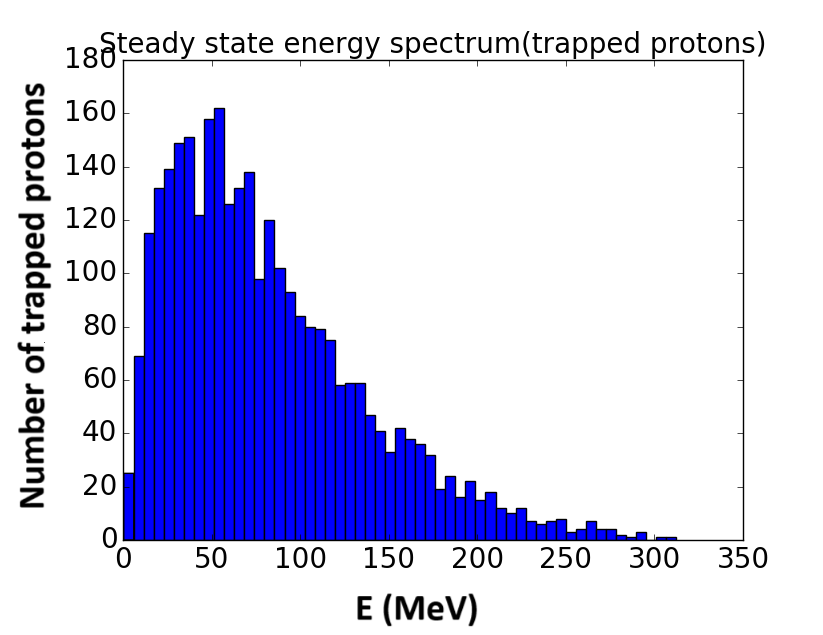}  
  \vspace{1em}
 \caption{ Energy spectrum of 4820 trapped protons at steady state in radiation belt. The $x$-axis represents energy of particles in MeV and the $y$-axis represents the number of particles in the corresponding energy bin.} 
  \label{fig:steadyE}
\end{figure}

\begin{figure}[h]
\centering
\includegraphics[width=0.8\linewidth]{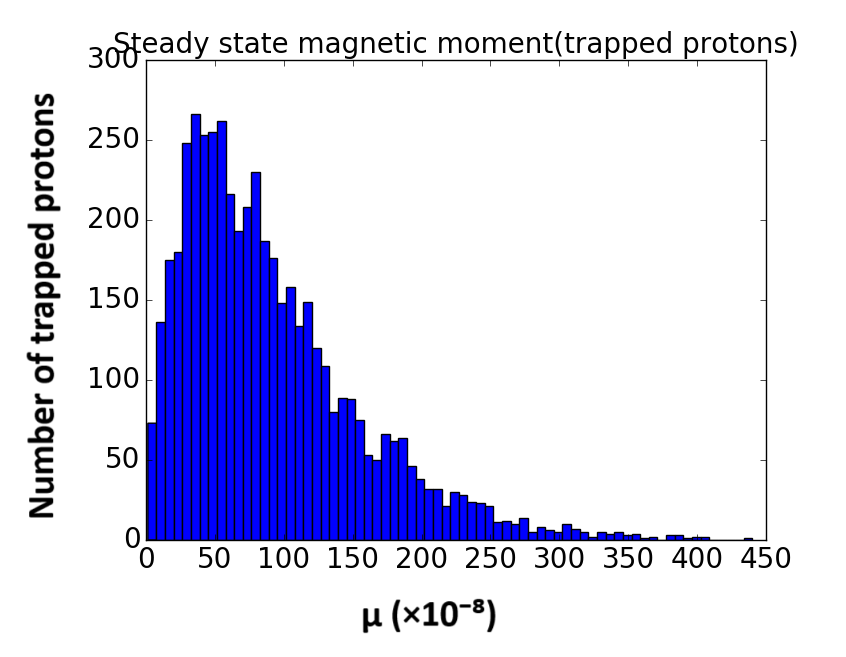}  
  \vspace{1em}
 \caption{ Magnetic moment distribution for 4820  trapped protons at steady state in radiation belt. The $x$-axis represents values of $\mu$. The $y$-axis represents the number of particles corresponding to each magnetic moment.} 
  \label{fig:steadymu}
\end{figure}

Figure~\ref{fig:steadyE} and Figure~\ref{fig:steadymu} represents the final steady-state energy and magnetic moment distributions respectively that has been obtained for an $\alpha$=10. These distribution functions obtained are similar to the initial seeded distribution.

The analytical method gives approximate particle guiding centre trajectory whereas the numerical simulation takes care of true gyromotion. Hence the loss cone could be properly studied using the current numerical simulations.

\textcolor{black}{We also comment on the chosen ensemble size of 10000 particles. In the present simulations, an initial ensemble of 10,000 particles evolves to a steady-state population of approximately 4820 trapped particles. The primary objective of this study is to quantify the overall trapped population and resulting precipitation statistics, for which resolving extremely fine-scale structure in the pitch-angle distribution near the loss-cone boundary is not critical. In this context, the retained ensemble size is adequate for the present analysis. 
This is further supported by the steady-state f(E–$\mu$) distribution shown in Fig.~\ref{fig:femu}, which exhibits a well-defined trapped population with a clear separation from the loss-cone region, without noticeable sparsity or irregular structure that would indicate strong shot-noise effects. The phase space is smoothly populated, indicating that the loss-cone boundary is sufficiently sampled and not determined by a small number of particles. While increasing the number of particles might further reduce finite-sampling effects, such calculations become computationally expensive for long-time integrations with full bounce and drift dynamics and are beyond the scope of the present study. Therefore, the ensemble size used here is sufficient for capturing the global trapped-particle dynamics and precipitation behavior considered in this work.}

\section{MHD Simulation of Alfv\'en wave packets}

Pre-seismic ULF/ELF waves are assumed to propagate along geomagnetic field lines in the inner Van Allen radiation belt in the form of Alfv\'en waves, a class of magnetohydrodynamic (MHD) waves. We consider two distinct types of wave inputs to represent different characteristics of seismo-electromagnetic emissions. Seismic emissions are often observed as narrowband enhancements superposed on broadband background noise. Accordingly, a narrowband coherent wave packet centered at selected ULF frequencies is used to model event-specific emissions and to examine whether they produce strong resonance-driven precipitation. In addition, a broadband incoherent spectrum, implemented as white noise with a flat power spectral density ($P(f)=\mathrm{constant}$), is included as a simplified, idealized stochastic representation of background emissions. This background is introduced to examine how its interaction with energetic particles compares with that of event-specific emissions, and whether it produces comparable precipitation or resonance-driven signatures. In this section, we describe these SEME-driven Alfvén wave inputs, including both narrowband coherent wave packets and broadband stochastic components, and their numerical implementation along the $L = 1.5$ shell.

The MHD equation describing the propagation of a dissipative Alfv\'en wave in a magnetized medium, driven by an external current source, is: 

\begin{equation}\label{alf_eqn}
 \frac{\partial ^ 2\bold v_1}  {\partial t ^2} -\bold {v_A}^2 \frac{\partial ^ 2\bold v_1} {\partial r ^2} +\frac{B_0} {\rho_{m0}} \frac{\partial J_{ext}}  {\partial t} - \eta \frac{\partial^2} { {\partial r}^2}  \frac{\partial \bold v_1}  {\partial t}  = 0
\end{equation}      

Here, $\bold{v_1}(r,t)$ represents a small perturbation of the medium velocity, $\bold{v_A}$ is the Alfv\'en velocity, $B_0$ is the background magnetic field strength, $\rho_{m0}$ is the background mass density, and $\eta$ is the kinematic viscosity of the magnetized medium. 

To solve Eq.~\ref{alf_eqn} numerically, we employ a one-dimensional Finite-Difference Time-Domain (FDTD) method. The Alfv\'en  velocity along the L-shell at $L = 1.5$ is calculated first. Since the Alfv\'en speed depends on both the magnetic field strength and medium's mass density, we first calculate the distributions of these quantities and then determine the resulting Alfv\'en velocity along the L-shell, as described in the following subsections.

\subsection{Geomagnetic field strength at $L = 1.5$}

%We calculate the magnetic field strength distribution along the L-shell at L = 1.5, which corresponds to the inner Van Allen radiation belts. The magnetic field strength along the L-shell at L=1.5 is calculated using the Earth's dipole model (see Eq.~\ref{mf_strength}) and is presented in Fig.~\ref{fig:mf}.

\begin{figure}[h]
\centering
  \includegraphics[width=0.8\linewidth]{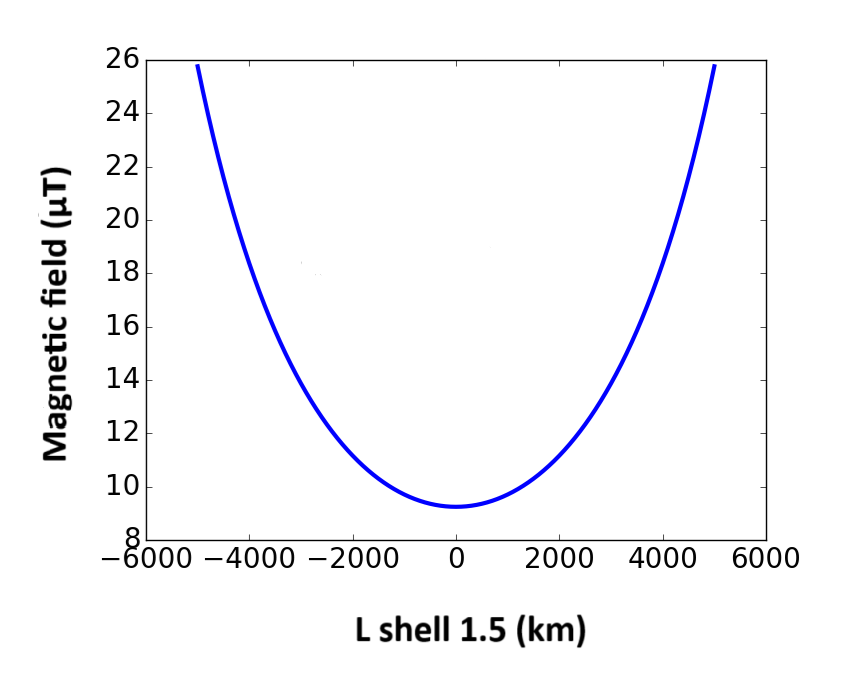}
  \vspace{1em}
 \caption{Magnetic field strength, $B(s)$ along the L = 1.5 shell in the inner Van Allen radiation belt, shown as a function of arc length $s$ along the field line. The field is calculated using the Earth's dipole model  (see Eq.~\ref{mf_strength}), yielding values of approximately $25\,\mu\mathrm{T}$ near the magnetic poles and $\sim 10\,\mu\mathrm{T}$ at the magnetic equator.}
  \label{fig:mf}
\end{figure}

The magnetic field strength distribution, $B(s)$ along the $L = 1.5$ shell, corresponding to the inner Van Allen radiation belt, is calculated using the Earth's dipole model (Eq.~\ref{mf_strength}). The resulting magnetic field strength distribution is shown in Fig.~\ref{fig:mf}.

%The Earth's dipole magnetic field strength along the L-shell at $L = 1.5$ attains a maximum value of $25\,\mu\text{T}$ near the magnetic poles and a minimum value of $10\,\mu\text{T}$ near the magnetic equator (see Fig.~\ref{fig:mf}).

Along this L-shell, the Earth's dipole magnetic field strength reaches a maximum of $25,\mu\text{T}$ near the magnetic poles and a minimum of $10,\mu\text{T}$ near the magnetic equator (see Fig.~\ref{fig:mf}).

\subsection{Mass density at $L = 1.5$}

\begin{figure}[h]
\centering
  \includegraphics[width=0.8\linewidth]{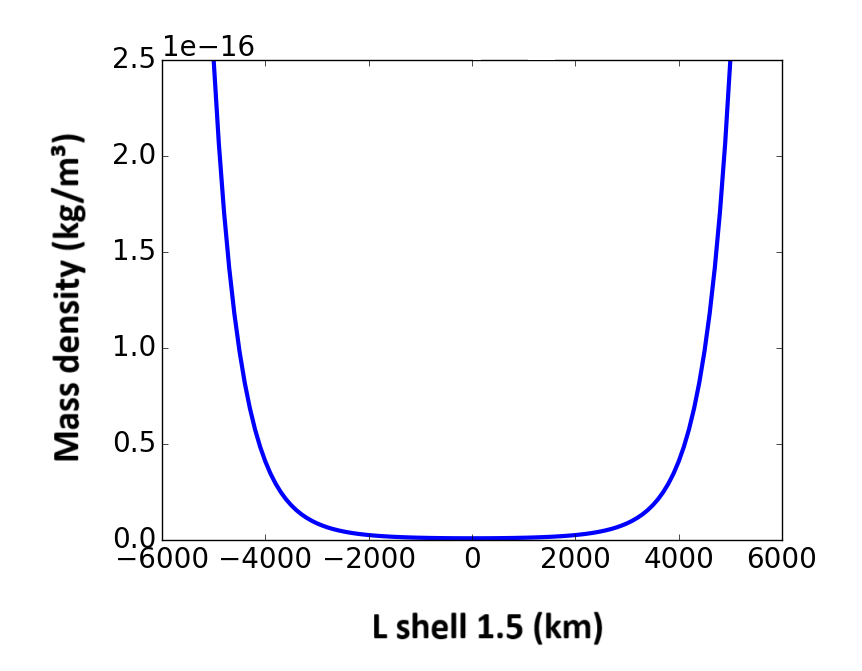}
  \vspace{1em}
 \caption{Mass density, $\rho_m(s)$ along the L = 1.5 shell in the inner Van Allen radiation belt, shown as a function of arc length $s$ along the field line.  The density decreases with increasing altitudes.}
  \label{fig:md}
\end{figure}

The  number density observed in the lower magnetosphere is given by Eq.~\ref{obs_chargedensity}.  The corresponding mass density as a function of altitude is expressed as:

\begin{equation}\label{eq:cd}
\rho_m(r)=n_0 {e^{-(r-R_I)/h}} \times 16 m_p +{ n_H(r)^{-1}} \times m_p
\end{equation}  
									
 where $m_p$ is the proton mass. The resulting mass density along $L$ shell 1.5  is presented in Fig.~\ref{fig:md}.

\subsection{Alfv\'en speed at $L = 1.5$}

The Alfv\'en wave speed at a point along the L-shell depends on the local magnetic field strength and the mass density, and is given by:

\begin{equation}\label{eq:cd}
\bold v_A(s)=\frac {B(s)}  {\sqrt{\mu_0 \rho_m(s)} }
\end{equation} 

\begin{figure}[h]
\centering
  \includegraphics[width=0.8\linewidth]{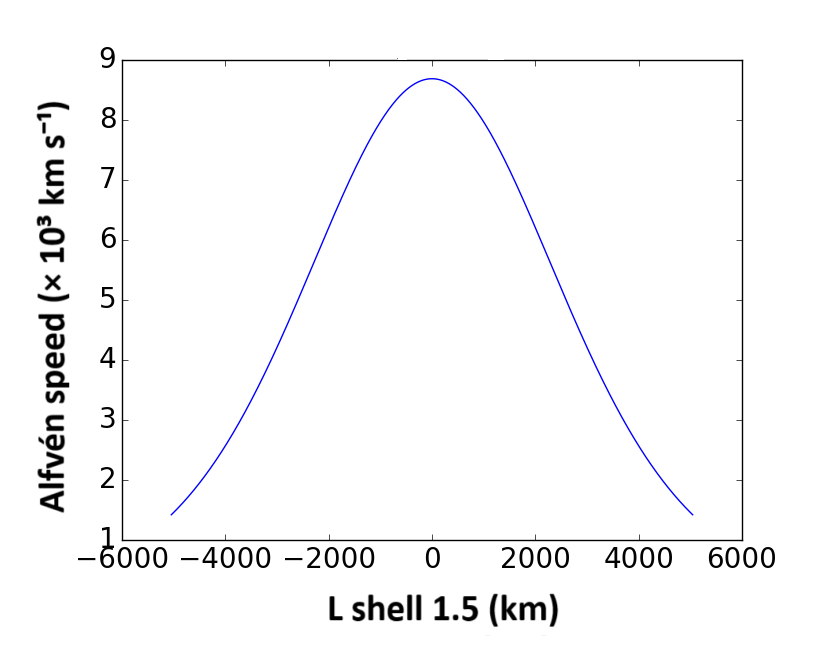}
  \vspace{1em}
 \caption{Alfvén speed, $\bold v_A(s)$ along the L = 1.5 shell in the inner Van Allen radiation belt, shown as a function of arc length $s$ along the field line. The wave propagation speed is higher near the magnetic equator and lower near the magnetic poles, indicating dispersive behavior along the $L$-shell.}
  \label{fig:as}
\end{figure}

The resulting Alfv\'en speed profile along the L-shell at $L=1.5$ is shown in Fig.~\ref{fig:as}.

As illustrated in Fig.~\ref{fig:as}, the nonuniform velocity profile causes Alfvén wave packets to experience dispersive propagation, traveling slower near the magnetic poles and faster near the equator.

\subsection{FDTD simulation of Alfv\'en wave packets}

We generate Alfv\'en wave packets along the L=1.5 shell by numerically solving the Alfv\'en wave equation (see Eq.~\ref{alf_eqn}) using a one-dimensional FDTD method. The scalar wave equation (Eq.~\ref{alf_eqn}) is discretized in space ($\delta s$) and time ($\delta t$) 
using a central difference scheme for numerical integration. The resulting discretized wave equation (Eq.~\ref{FDTD1}) is:

\begin{equation}\label{FDTD1}
\begin{aligned}
{\bold {u_i}^{n+1}(1+P)}=2\bold {u_i}^{n}-\bold {u_i}^{n-1}+S ^2 ({\bold {u_{i+1}}^{n}-2\bold {u_i}^{n}+\bold {u_{i-1}}^{n}})+\\ \frac{P} {2} ({ u_{i+1}^{n+1}  +u_{i-1}^{n+1}+u_{i+1}^{n-1} -2u_{i}^{n-1}-u_{i-1}^{n-1}})- \\
\frac {B} { \rho_{m0}}\frac  { \bold J_i^{n+\frac {1} {2}} -\bold J_i^{n-\frac{1}{2}} } {2\delta t} 
\end{aligned}
\end{equation}

where, 

\begin{equation}\label{FDTD2}
%\begin{aligned}
S= \frac {\bold v_A \delta t} {\delta s} ;
P=\frac {\eta \delta t}  {(\delta s)^2}
%\end{aligned}
\end{equation}

The numerical dispersion in our FDTD simulation is minimized by choosing a high temporal resolution of $T / \delta t = 100$, 
where $T$ is the wave period and $\delta t$ is the discrete time step. This choice also satisfies the Nyquist criterion. 
For the spatial mesh, the Courant condition is applied, which requires that the Courant number $S$ of the generated wave satisfy $S \le 1$ for stability. 
Our choice of $S = 1$ ensures stable wave propagation. The damping parameter $P$ is set to zero, corresponding to a lossless wave. 
Various nonzero values of $P$ were tested, but realistic values of $\eta$ were too small to produce a noticeable effect. 
The Alfv\'en wave packet is driven by the external current source given in Eq.~~\ref{J_cs}, which generates a narrowband Gaussian Alfv\'en wave packet.

\begin{equation}\label{J_cs}
J= J_0cos(2\pi ft ) e^{   (\frac{s-s_0}  { \sigma_w})   ^2 } e^{   (\frac{t-t_0}  { \sigma_t})   ^2 }
\end{equation}

In the present study, we numerically solve for both narrowband Alfv\'en wave packets at various frequencies and broadband Alfv\'en wave packets representing a background noisy spectrum along the L-shell at $L=1.5$. These two types of waves allow us to distinguish between wave–particle interactions during event-specific interactions and general background interactions.

The parameters of the current source used to generate a narrowband Alfvén wave packet are chosen as follows:

\begin{equation}\label{eq:Jp}
\begin{aligned}
J_0    =  10 ^ {-18}; 
f = 30\, \text{Hz} ;
\sigma_w     =  200\,  \text{kms}; \\
s_0 = 0\, \text{km} ; 
\sigma_t     = 2  \frac{1}{f} ; 
t_0      = 10 \frac{1}{f}
\end{aligned}
\end{equation}

% A narrow band Alfv\'en wave packet with a fixed central frequency at 30 Hz is simulated and a time instant of this monochromatic Alfv\'en wave  is presented in Fig.~\ref{fig:Alf_timesnap}.

\begin{figure}[h]
\centering
  \includegraphics[width=0.8\linewidth]{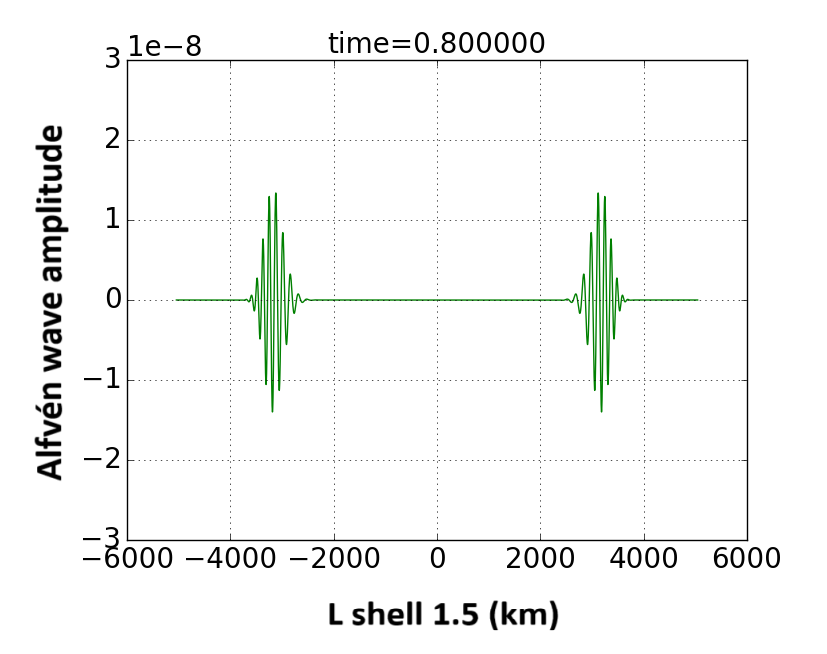}
  \vspace{1em}
 \caption{ Time snap of a propagating narrowband Gaussian Alfv\'en packet along the  L=1.5 shell, with a fixed central frequency of 30 Hz, generated by the current source (see Eqs.~\ref{J_cs}-\ref{eq:Jp}) The simulation is performed using FDTD methods by discretizing the scalar Alfv\'en wave equation (see Eqs.~\ref{FDTD1}-\ref{FDTD2}).}
  \label{fig:Alf_timesnap}
\end{figure}

 We present in Fig.~\ref{fig:Alf_timesnap} a time snapshot of a propagating monochromatic Alfvén wave, generated from our FDTD MHD simulations of a narrowband Alfvén wave packet with a fixed central frequency of 30 Hz.
						
\textcolor{black}{In the present model, the source amplitude $J_0 = 10^{-18}$ is introduced phenomenologically to generate a weak Alfv\'enic perturbation for the wave--particle interaction study, rather than being derived from a self-consistent seismic source model. The current value of $J_0$ from Eq.~\ref{eq:Jp} specifies the external current driver (Eq.~\ref{J_cs}), while the magnetic perturbation amplitude at L-shell 1.5 is the response of the inhomogeneous MHD system obtained after solving the FDTD wave equation (Eq.~\ref{alf_eqn}). This value of $J_0$ is therefore chosen to obtain a corresponding magnetic perturbation amplitude used in the particle-interaction stage of order $B_{\rm wave} \sim 10^{-8}\,{\rm T}$ (about 10 nT). Along the modeled $L = 1.5$ shell, where the background field is $B_0 \sim 10\text{--}25\,\mu{\rm T}$, this corresponds to $\delta B/B_0 \sim 4\times10^{-4}\text{--}10^{-3}$, i.e., a perturbation at the $0.04\%\text{--}0.1\%$ level. Thus, the wave remains a small perturbation relative to the background field. In comparison with reported pre-seismic ULF magnetic anomalies, which report fluctuations close to the sub-nT to few-nT range\citep{Alimoradi2024}, our adopted amplitude is slightly larger but remains within a broader, plausible nT-scale range that may reasonably be expected. However, the current choice is for the present proof-of-principle study and is used here to test whether such weak Alfv\'enic disturbances can produce measurable resonant precipitation. A self-consistent derivation of $J_0$ from observed seismo-electromagnetic source strengths, including transmission losses through the atmosphere and ionosphere, is not addressed in the present work.}

\begin{figure}[h]
\centering
  \includegraphics[width=0.8\linewidth]{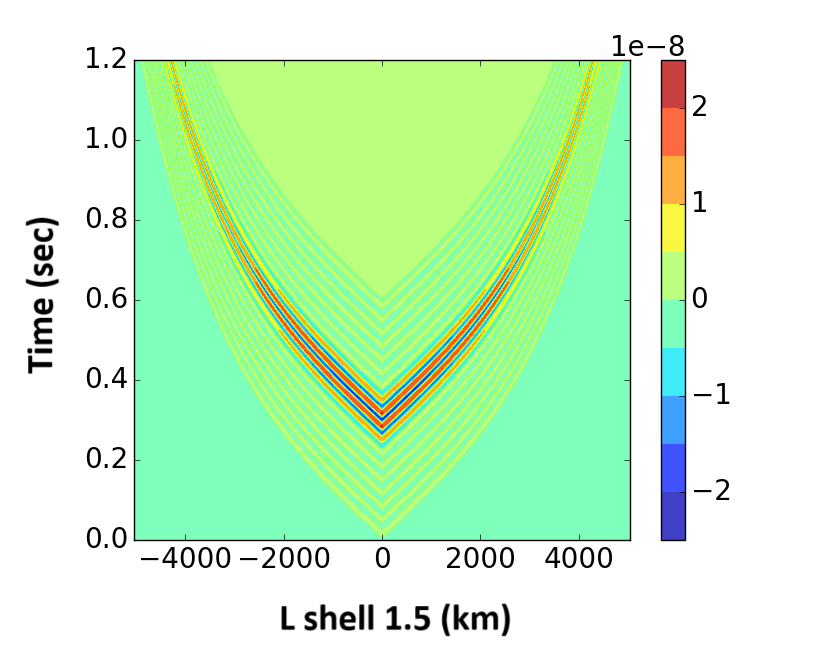}
  \vspace{1em}
 \caption{Contour plot of the propagating narrowband Gaussian Alfv\'en wave packet shown in Fig.~\ref{fig:Alf_timesnap} along the L = 1.5 shell, with a fixed central frequency of 30~Hz, generated by the current source (see Eqs.~\ref{J_cs}-\ref{eq:Jp}). The wave slows down as it propagates toward the magnetic poles and is squeezed, which appears as bending contours in the figure. The bright and dark fringes indicate the 30~Hz oscillations of the fluctuating field.
}
  \label{fig:Alf_contourplot}
\end{figure}

Fig.~\ref{fig:Alf_contourplot} presents the contour plot of the Alfvén wave packet at a central frequency of 30 Hz, depicted in Fig.~\ref{fig:Alf_timesnap}. \textcolor{black}{Along the dipolar $L = 1.5$ field line, the Alfv\'en speed decreases toward the polar regions compared to the equator, resulting in a nonuniform propagation medium along the field line. Consequently, the Alfv\'en wave packet propagates more slowly near the poles and more rapidly near the equator (see Fig.~\ref{fig:as}). As the wave approaches the polar region, the reduced propagation speed leads to a compression of the wave packet (see Fig.~\ref{fig:Alf_timesnap}), producing the apparent ``squeezing'' or bending of the contour structure observed in Fig.~\ref{fig:Alf_contourplot}. Such behavior is expected for wave propagation in a dipolar magnetic geometry with spatially varying Alfv\'en speed. The squeezing observed in Figs.~10–12 is not caused by numerical discretization or grid-spacing effects in the FDTD scheme. Numerical dispersion is minimized by employing high temporal resolution ($T/\delta t = 100$), satisfying the Nyquist criterion, while stability is ensured by enforcing the Courant condition with $S \leq 1$, with $S = 1$ chosen for stable wave propagation. Rather, the observed behavior reflects the underlying physics of wave propagation in a nonuniform magnetic field with a spatially varying Alfv\'en-speed profile along the field line.}
						
\begin{figure}[h]
\centering
  \includegraphics[width=0.8\linewidth]{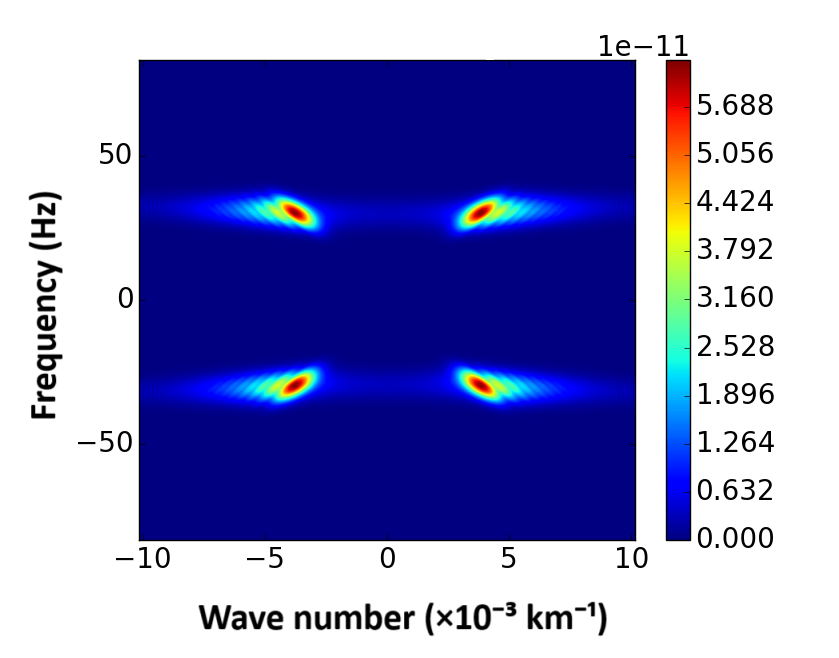}
  \vspace{1em}
 \caption{Dispersion plot of the propagating narrowband Gaussian Alfv\'en wave packet (see Fig.~\ref{fig:Alf_timesnap}) in $\omega$--$k$ space. 
The plot shows the relationship between the angular frequency $\omega$ and the wave number $k$, highlighting the dispersive nature of the wave due to the nonuniform Alfv\'en speed along the L-shell. 
The bright regions correspond to dominant spectral components of the wave packet. }
  \label{fig:Alf_disp}
\end{figure}

Fig.~\ref{fig:Alf_disp} presents the dispersion plot of the Alfvén wave packet depicted in Fig.~\ref{fig:Alf_timesnap}.

It corresponds to the same wave packets being inserted again and again, and the result after transforming both space and time. The resulting wave is spread out in $k$ but still narrow in $\omega$.

\begin{figure}[h]
\centering
  \includegraphics[width=0.8\linewidth]{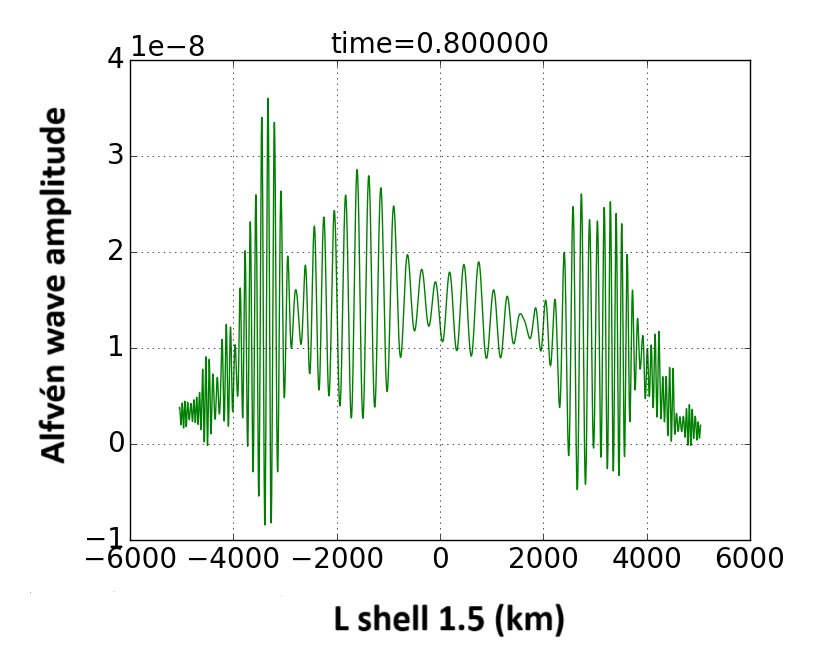}
  \vspace{1em}
 \caption{Time snapshot of a propagating broadband noisy Alfvén field along the L = 1.5 shell. The wave is composed of multiple low-intensity wave packets, each with randomly chosen frequency, amplitude, spatial width, temporal width, and position, varied around the parameters of Eq.~\ref{eq:Jp}, with frequencies uniformly distributed between 5 Hz and 60 Hz. The simulation is performed using FDTD methods by discretizing the scalar Alfv\'en wave equation (see Eqs.~\ref{FDTD1}-\ref{FDTD2}).}
  \label{fig:Alfnoise_time}
\end{figure}

\begin{figure}[h]
\centering
  \includegraphics[width=0.8\linewidth]{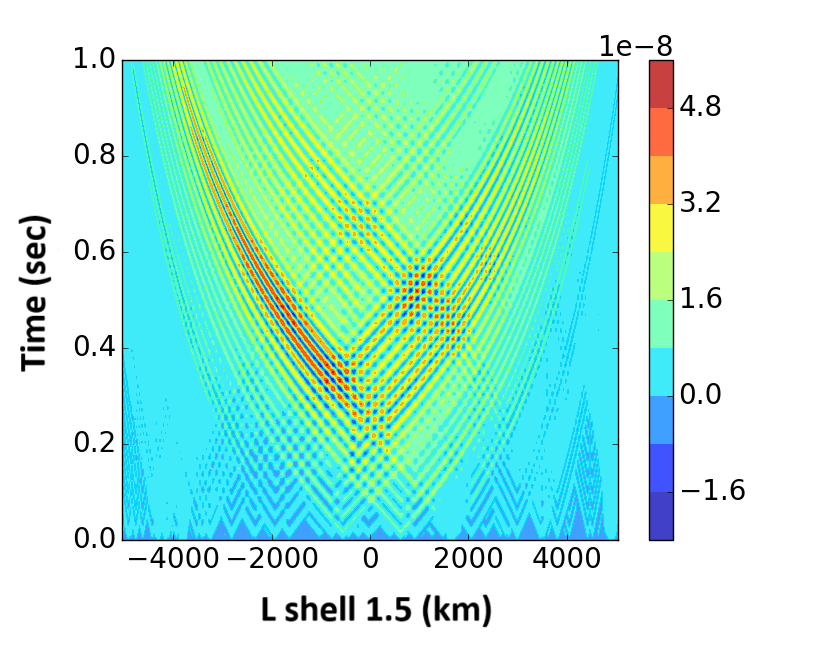}
  \vspace{1em}
 \caption{Contour plot of the propagating broadband Alfv\'en Noise (see Fig.~\ref{fig:Alfnoise_time}) along the L-shell 1.5. The wave is composed of multiple low-intensity wave packets, each with randomly chosen frequencies, amplitudes, spatial widths, temporal widths, and positions, with frequencies uniformly distributed between 5 and 60 Hz. The packets are distributed randomly in space and time, producing a complex, irregular pattern of fluctuating fields. The simulation is performed using FDTD methods by discretizing the scalar Alfv\'en wave equation (see Eqs.~\ref{FDTD1}-\ref{FDTD2}).}
  \label{fig:Alfnoise_cp}
\end{figure}

\begin{figure}[h]
\centering
  \includegraphics[width=0.8\linewidth]{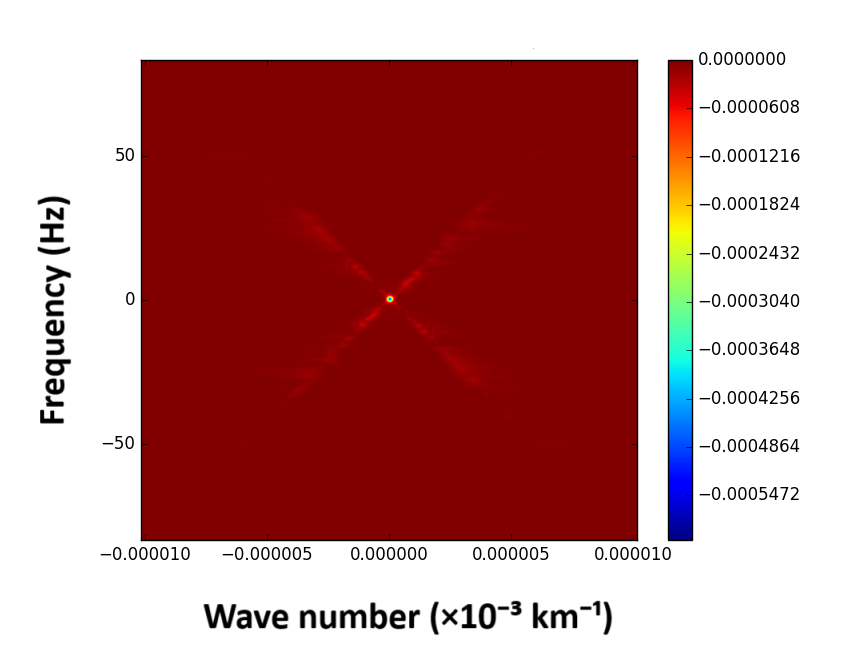}
  \vspace{1em}
 \caption{Dispersion plot of the Alfv\'en Noise (see Fig.~\ref{fig:Alfnoise_time}).}
  \label{fig:Alfnoise_disp}
\end{figure}

Next, we present FDTD simulations of a broadband noisy Alfvén wave using Eq.~\ref{FDTD1}, shown in Fig.~\ref{fig:Alfnoise_time}, Fig.~\ref{fig:Alfnoise_cp}, and Fig.~\ref{fig:Alfnoise_disp}. 
\textcolor{black}{This broadband Alfvén noise is introduced as a simplified, idealized background model composed of multiple low-intensity wave packets with randomly chosen positions, times, amplitudes, and spatial/temporal widths, with frequencies uniformly distributed between 5 and 60 Hz. The packet parameters were varied around the values of Eq.~\ref{eq:Jp} to generate a representative broadband noisy signal. This construction corresponds to an approximately white-noise spectrum with an approximately flat power spectral density and should be regarded as an idealized broadband representation. The purpose of this model is to test whether a representative broadband background with a comparable overall fluctuation level—chosen to have the same RMS amplitude as the monochromatic narrowband wave packets—can produce precipitation comparable to the resonant narrowband case, rather than to reproduce the exact magnetospheric power spectral density.} A time instant of the noisy Alfv\'en wave is presented in Fig.~\ref{fig:Alfnoise_time}.

%The contour plot of the  Alfv\'en noise in Fig.~\ref{fig:Alfnoise_time} is presented in Fig.~\ref{fig:Alfnoise_cp}.

Fig.~\ref{fig:Alfnoise_cp} presents the contour plot of the Alfv\'en noise depicted in Fig.~\ref{fig:Alfnoise_time}.

%The dispersion diagram of the broadband Alfv\'en noise in Fig.~\ref{fig:Alfnoise_time} is presented below in Fig.~\ref{fig:Alfnoise_disp}.

Fig.~\ref{fig:Alfnoise_disp} presents the dispersion diagram of the broadband Alfv\'en noise depicted in Fig.~\ref{fig:Alfnoise_time}.

The dispersion of a 30Hz Alfv\'en wave is for a single wave packet with fixed single central parameters whereas the noisy Alfv\'en wave is a combination of all kinds of wave packets with different frequencies, amplitudes, pulse widths and initial positions and times. In Fig.~\ref{fig:Alfnoise_cp} this is seen as a linear superposition of all these wave packets, uniformly generated along the L shell and at different instances of time. The difference between the single wave packet and multiple wave packets are visible in their contour plots. In Fig.~\ref{fig:Alfnoise_disp} the waves spread out in $k$ space for different frequencies and is seen in this plot. A single wave packet is not distinct here in Fig.~\ref{fig:Alfnoise_disp} compared to Fig.~\ref{fig:Alf_disp} since there are multiple wave packets superimposed. Also since the point $(k,\omega) = (0,0)$ has not been removed from the plot, the contrast in the color bar is less in this current plot for noisy Alfv\'en wave.

\section{Interaction of Alfv\'en wave packets with energetic particles in inner Van Allen belt}

%The study of energetic particle interaction with Alfvén wave packets in the inner Van Allen belt using numerical simulations is presented here. A kinetic model of the 3D steady-state energetic particle population, approximately resembling the density profile of the inner radiation belt, was developed in Section 2. Alfv\'en wave generated along the 1D geomagnetic line or '$s$' coordinate along L=1.5 was presented in Section 3. This Alfv\'en wave in 1D polar '$s$' coordinate system is converted to 3D Cartesian coordinate system below where it's amplitude is in the direction parallel to the plane of the geomagnetic line ie in the $x-z$ direction.  The magnetic field intensity of the Alfv\'en wave is absent in the direction perpendicular to this plane. $B_{As}$ is the Alfv\'en wave component along the 1D '$s$' coordinate and $B_{Ax}$, $B_{Ay}$ and $B_{Az}$ are the magnetic field components in the Cartesian coordinates. 

The study of steady state trapped energetic proton interactions with Alfvén wave packets in the inner Van Allen belt using numerical simulations is presented here. In Section 2, we developed a kinetic model of the 3D steady-state energetic particle population, designed to approximately reproduce the density profile of the inner radiation belt. In Section 3, the Alfvén wave disturbance generated along the 1D geomagnetic field line (the '$s$' coordinate) at L = 1.5 was described. This 1D Alfvén wave, defined along the polar '$s$' coordinate before, is now converted to a 3D Cartesian coordinate system, where its amplitude lies in the plane of the geomagnetic line (the $x$–$z$ plane). The wave has no magnetic field component perpendicular to this plane. Here, $B_{As}$ denotes the Alfvén wave component along the 1D '$s$' coordinate, while $B_{Ax}$, $B_{Ay}$, and $B_{Az}$ correspond to the magnetic field components in the Cartesian system.

\begin{equation}
%\begin{aligned}
   B_{Ax} = B_{As} cos\theta ;
   B_{Ay} = 0 ;
  B_{Az} =B_{As} sin \theta
%\end{aligned}
\end{equation}

 %The particle trajectories are now obtained from the Lorentz force equation, Eq.~\ref{LF} and is solved using the same RK4 algorithm.  The presence of the Alfv\'en disturbance is superimposed on the background static geomagnetic field of the Earth which gives the total magnetic field in the Lorentz force equation.  

 The particle trajectories are then computed using the Lorentz force equation (Eq.~\ref{LF}), which is solved with the same RK4 integration algorithm. The total magnetic field in the Lorentz force equation is now obtained by superimposing the Alfvén wave disturbance on the background static geomagnetic field of the Earth.

 \begin{equation}\label{eq:dxvst}
\begin{aligned}
   {B_{total}}= {B_{static}} + {B_{Alfven}}
 \end{aligned}
\end{equation}

 When a particle moves along the field lines in the trajectory solver, it experiences both the background geomagnetic field and the superimposed Alfv\'en disturbance. Consequently, the Lorentz force in the presence of the Alfv\'en disturbance is expressed as:

\begin{equation}\label{LF2}
\begin{aligned}
   m \frac{dv_x}{dt}=q(v_y(B_{sz}+B_{Az}) -v_z(B_{sy} +B_{Ay})) \\
  m \frac{dv_y}{dt}=q(v_z(B_{sx}+B_{Ax}) -v_x(B_{sz} +B_{Az})) \\
  m \frac{dv_z}{dt}=q(v_x(B_{sy}+B_{Ay}) -v_y(B_{sx} +B_{Ax}))
\end{aligned}
\end{equation}

The steady-state energetic particle population (4820 particles, after removing those lost in the loss cone) is evolved using Eq.~\ref{LF2}, with Alfvén waves superimposed on the background geomagnetic field. Particle trajectories are computed for 1.5 s. During the first 1 s, wave–particle interactions occur in the presence of both the background geomagnetic field and the imposed Alfv\'en wave. In the remaining 0.5 s, the Alfv\'en wave is switched off, and particle dynamics evolve under the background geomagnetic field alone, allowing us to isolate the net effect of the wave–particle interaction and examine the subsequent particle evolution. The bounce periods of particles across different energies range from 0.1 to 0.6 s, ensuring that each particle undergoes multiple bounces within the interaction time. As a result of the interaction, particle trajectories are modified due to pitch-angle scattering. Particles that fall below the mirror point at 1000 km (entering the Earth’s ionosphere or atmosphere) are frozen in the simulation and counted as lost.

\subsection{Count of energetic particle precipitation due to interaction with  different Alfv\'en wave packets}

\subsubsection{Narrowband Alfv\'en wave packets}

\begin{figure}[h]
\centering
  \includegraphics[width=0.8\linewidth]{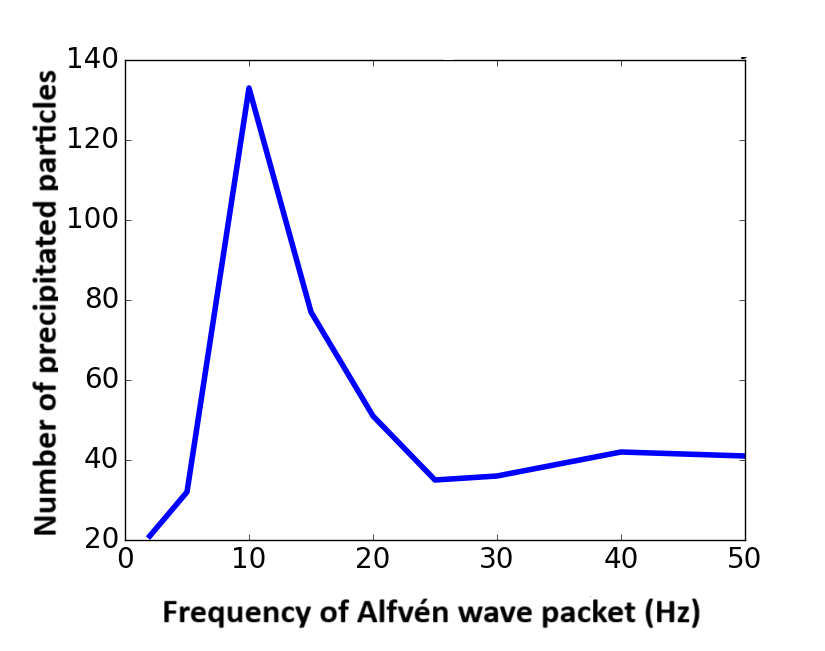}
   \vspace{1em}
 \caption{Number of particles precipitated as a function of the central frequency of a narrowband Gaussian Alfvén wave packet. The precipitation peak at 10 Hz is attributed to resonance condition.}
  \label{fig:partprec_freq}
\end{figure}

We present the results of the interaction between the steady-state trapped energetic particle population (see Fig.~\ref{fig:density_alpha10}) and Alfv\'en disturbances in the form of narrowband Alfv\'en wave packets with a fixed central frequency (see Fig.~\ref{fig:Alf_timesnap} for a 30 Hz wave packet). In different wave–particle interaction simulations, the central frequency of the narrowband Alfv\'en wave packets was varied (e.g., 2, 5, 10, 20, 30, 40, and 50 Hz), and the number of particles precipitated in each case is shown in Fig.~\ref{fig:partprec_freq}.

The results indicate enhanced energetic particle precipitation at a central frequency of 10 Hz, consistent with a resonance condition between energetic protons and Alfvén waves in the inner radiation belt. The presence of the Alfvén disturbance induces pitch-angle scattering, leading to enhanced particle precipitation from the belt. As shown later in Section 4.2, this resonance arises from cyclotron resonance (n=1) between oppositely directed wave and particle motion.

\subsubsection{Broadband Alfv\'en wave packet}

In general, background Alfvén noise is present at all times and can interact with trapped energetic particles in the inner radiation belt. To assess the relative contribution of background noise compared to a strong seismic or resonant disturbance, we simulate the interaction of broadband Alfvén noise with the steady-state energetic particle population. This allows us to distinguish particle detrapping driven by resonant events from that produced by the usual background fluctuations. \textcolor{black}{The broadband Alfvén noise considered here is constructed as an simplified idealized background composed of multiple low-intensity wave packets with randomly chosen parameters and frequencies uniformly distributed between 5 and 60 Hz, corresponding to an approximately white-noise spectrum with a constant or nearly flat power spectral density (P(f)=constant). This simplified model is intended as an idealized representative broadband benchmark rather than an exact representation of magnetospheric fluctuations. For comparison, the broadband Alfvén noise is generated with the same RMS amplitude as the monochromatic narrowband Alfvén wave packets used in the previous simulations, enabling a direct comparison of particle precipitation under comparable fluctuation levels. Although real magnetospheric background noise may exhibit non-uniform or 1/f-like spectra, the present white-noise model provides a baseline for assessing whether broadband background fluctuations alone can produce precipitation comparable to the resonant narrowband case.}

Table~\ref{tab:noisevsres} summarizes the number of particles precipitated due to resonant narrowband Alfv\'en wave packet, non-resonant narrowband Alfve\'n wave packet and background broadband Alfv\'en noise.

\begin{table}[htbp]
  \caption{Particle precipitation for resonant narrowband, non-resonant narrowband, and broadband Alfvén disturbances}
  \begin{center}
  \begin{tabular}[c]{|c|r|} \hline
  
   Alfvén wave disturbance  &  Particles precipitated  \\ \hline
   Resonant narrowband packet &  133   \\
   Non-Resonant narrowband packet &  41   \\
   Broadband Alfv\'en noise &  36  \\ \hline
    
  \end{tabular}
  \label{tab:noisevsres}
  \end{center}
\end{table}

The results of simulations of the interaction between energetic trapped particles (see Fig.~\ref{fig:density_alpha10}) and broadband Alfv'en noise (see Fig.~\ref{fig:Alfnoise_time}) show that only 36 particles are precipitated or detrapped due to the background broadband noise. This is comparable to the non-resonant narrowband case, where 41 particles are precipitated, and significantly smaller than the resonant narrowband case, where 133 particles are precipitated, indicating that resonant interactions play a dominant role in enhanced particle precipitation.

\subsection{Energy spectrum and pitch angle scattering of precipitated particles }

\subsubsection{Narrowband Alfv\'en wave packet}

\begin{figure}[h]
\centering
  \includegraphics[width=0.8\linewidth]{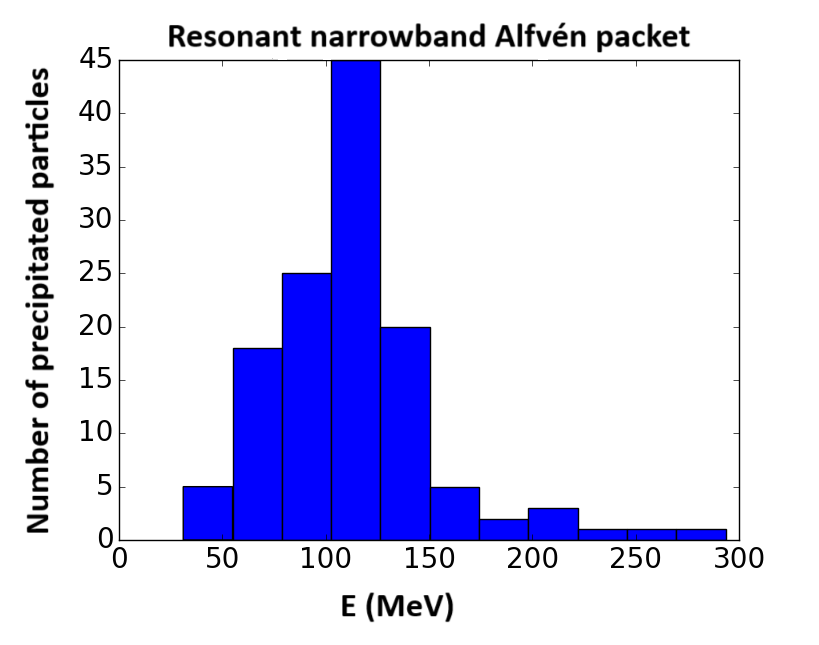}
  \vspace{1em}
 \caption{Energy spectrum of particles precipitated due to resonant narrowband Alfvén wave interaction. The precipitated particles lie predominantly in the range 100–130 MeV, with a peak at 125 MeV.} \label{fig:energyspectrum_resonance}
  \label{fig:boat1}
\end{figure}

The energy spectrum of energetic protons precipitated from the radiation belt due to interaction with Alfv\'en waves at the resonant frequency of 10 Hz is presented in Fig.~\ref{fig:energyspectrum_resonance}.

The results indicate that most particle losses occur in the 100–130 MeV range, with a peak around 125 MeV, indicating a selective interaction near 125 MeV, suggestive of resonance with the Alfvén wave.

\begin{figure}[h]
\centering
  \includegraphics[width=0.8\linewidth]{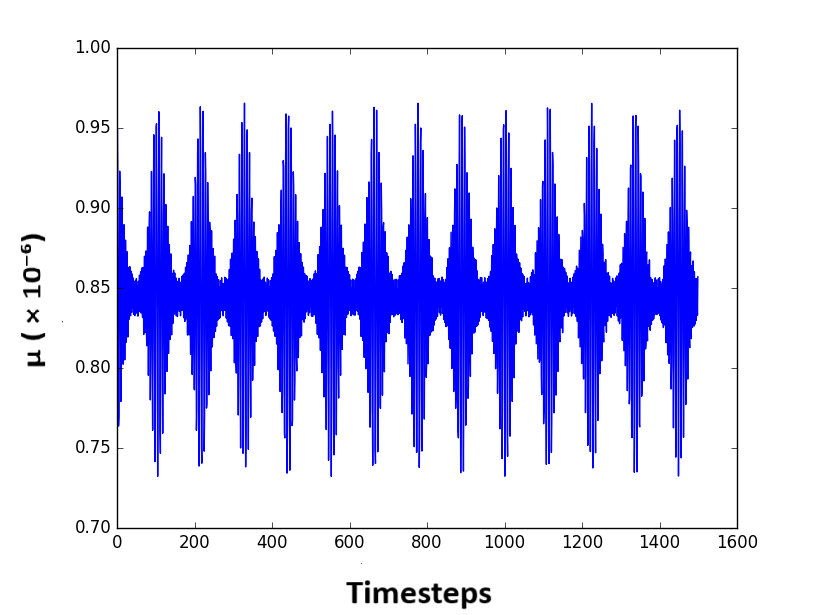}
  \vspace{1em}
 \caption{Magnetic moment of 125 MeV particle before interaction with a resonant narrowband Alfv\'en wave of 10 Hz. The particle’s magnetic moment fluctuates during individual gyro-motions but remains constant on average over multiple bounce periods, consistent with adiabatic invariance.}
  \label{fig:pa_beforeint}
\end{figure}

 The magnetic moment of a 125 MeV particle before interaction with a 10 Hz Alfv\'en wave packet is presented in Fig.~\ref{fig:pa_beforeint}. The particle undergoes gyro-motion and bounce motion along the Earth's magnetic field lines. While the magnetic moment varies during individual gyro motions, its average value remains constant over bounce periods, consistent with adiabatic invariance.

\begin{figure}[h]
\centering
  \includegraphics[width=0.8\linewidth]{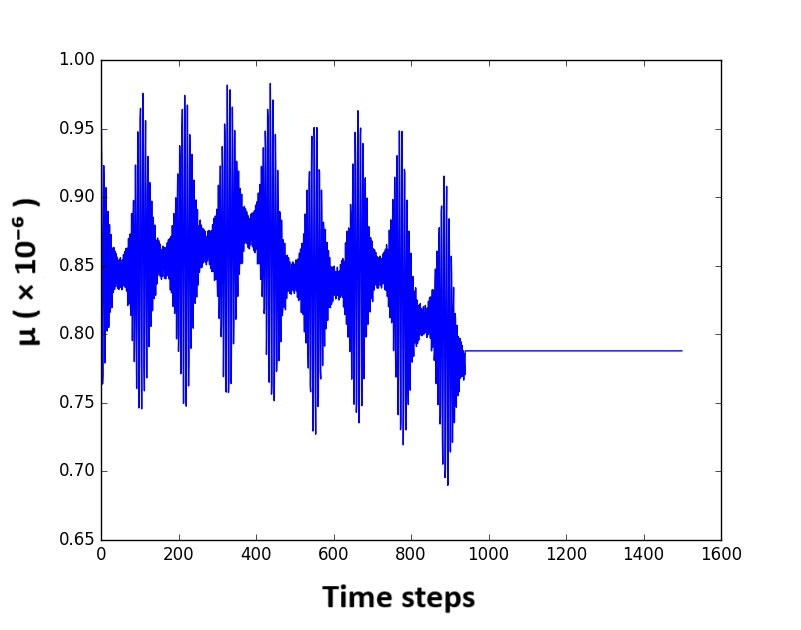}
  \vspace{1em}
 \caption{Magnetic moment of a 125 MeV particle after interaction with a resonant narrowband Alfv\'en wave. The particle undergoes scattering in $\mu$ reflecting pitch-angle changes induced by the Alfvén wave, until it becomes untrapped and precipitates. The change in average magnetic moment occurs is seen between time steps 400 and 900. After precipitation, the particle is frozen in the simulation, with its final magnetic moment indicated by the thin blue line.}
  \label{fig:pa_afterint}
\end{figure}

The magnetic moment of the same 125 MeV particle after interaction with a 10 Hz Alfv\'en wave packet is presented in Fig.~\ref{fig:pa_afterint}.

After interacting with the 10 Hz resonant Alfv\'en wave packet, the particle is scattered in $\mu$ space, indicating that the first adiabatic invariant is broken, ultimately becoming untrapped and precipitated. The change in average magnetic moment occurs primarily between time steps 400 and 900, reflecting pitch-angle scattering and particle detrapping induced by the Alfvén wave. After 900 time steps, the particle is precipitated and frozen in the simulation, represented by the thin blue line, which remains constant to indicate the particle’s final magnetic moment.  

\textcolor{black}{To verify that this interaction corresponds to cyclotron resonance, we examine the Doppler-shifted cyclotron resonance condition for oppositely directed wave and particle motion ($n=1$):
\begin{equation}
\omega - k_\parallel v_\parallel = n\Omega_c
\end{equation}}

\textcolor{black}{A representative quantitative check of the cyclotron resonance condition may be carried out for the 125 MeV proton with average magnetic moment $\mu \approx 0.85 \times 10^{-6}$, shown in Fig.~\ref{fig:pa_afterint}, at the stage where its magnetic moment begins to change significantly. We take this stage as the onset of strong wave--particle interaction and evaluate the local background magnetic field and corresponding Alfv\'en speed at the particle position along the $L = 1.5$ shell. For a 125 MeV proton, the total speed is $v \approx 1.58 \times 10^8\,\mathrm{m\,s^{-1}}$. At this location, the local magnetic field is $B = 16\,\mu\mathrm{T}$, and the corresponding Alfv\'en speed is $V_A \approx 3.66 \times 10^6\,\mathrm{m\,s^{-1}}$. Using the simulated magnetic moment, we obtain $v_\perp \approx 1.30 \times 10^8\,\mathrm{m\,s^{-1}}$ and $v_\parallel \approx 8.94 \times 10^7\,\mathrm{m\,s^{-1}}$.}

\textcolor{black}{For the 10 Hz wave, $\omega = 2\pi f \approx 62.8\,\mathrm{rad\,s^{-1}}$. Using the cyclotron resonance condition for oppositely directed wave and particle motion, $\omega + k_\parallel v_\parallel = \Omega_c$, together with $k_\parallel = \omega/V_A$, we obtain $k_\parallel \approx 1.72 \times 10^{-5}\,\mathrm{m^{-1}}$. This gives $\omega + k_\parallel v_\parallel \approx 62.8 + 1537 \approx 1.60 \times 10^3\,\mathrm{rad\,s^{-1}}$, in agreement with $\Omega_c = qB/m \approx 1.60 \times 10^3\,\mathrm{rad\,s^{-1}}$. Thus, the 10 Hz wave satisfies the cyclotron resonance condition for $\sim 125$ MeV protons for oppositely directed wave and particle motion in the present simulations.}

%The average $\mu$ of the particle decreases after interaction. The Alfv\'en wave amplitude in our simulation has an amplitude 1e-8. Where as the change of $\mu$ is of the order of 1e-7. Therefore even though disturbance increases B, the overall $\mu$ decreases and hence $v_\perp$ decreases as seen from the equation,

%\begin{equation}\label{eq:dxvst}
%v_\perp= \sqrt{\frac{2 \mu B}{m}}
%\end{equation}

% As a result the parallel speed increases and particle's bounce motion is increased such that it reaches the lower magnetosphere boundary. This is how a particle gets detrapped and lost. 

Although the Alfvén wave has a small amplitude 
($B_{\rm wave} \sim 10^{-8}$), the change in $\mu$ is of the order $10^{-7}$. 
Consequently, despite the local increase in magnetic field due to the disturbance, 
the particle’s average magnetic moment decreases, leading to a reduction in the 
perpendicular velocity:

\begin{equation}\label{eq:dxvst}
v_\perp = \sqrt{\frac{2 \mu B}{m}}
\end{equation}

As $v_\perp$ decreases, the parallel velocity increases, enhancing the particle’s 
bounce motion until it reaches the lower magnetosphere boundary. 
This mechanism explains how the particle becomes detrapped and lost from the radiation belt.

\textcolor{black}{Additionally, we estimate the implied pitch-angle diffusion coefficient, $D_{\alpha\alpha}$, based on the change of $\mu$ shown for the wave–particle resonance in Fig.~\ref{fig:pa_afterint}, and compare it with quasi-linear theory (QLT), which provides additional quantitative insight into the resonant wave–particle interaction physics investigated in this work. The calculation of the diffusion coefficient from our simulations and the comparison with analytic QLT are presented in Appendix~\ref{AppenB1}. We find that the diffusion implied by the simulation is broadly consistent, to within an order of magnitude, with the
quasi-linear expectation for weak resonant scattering.} 

\begin{figure}[h]
\centering
  \includegraphics[width=0.8\linewidth]{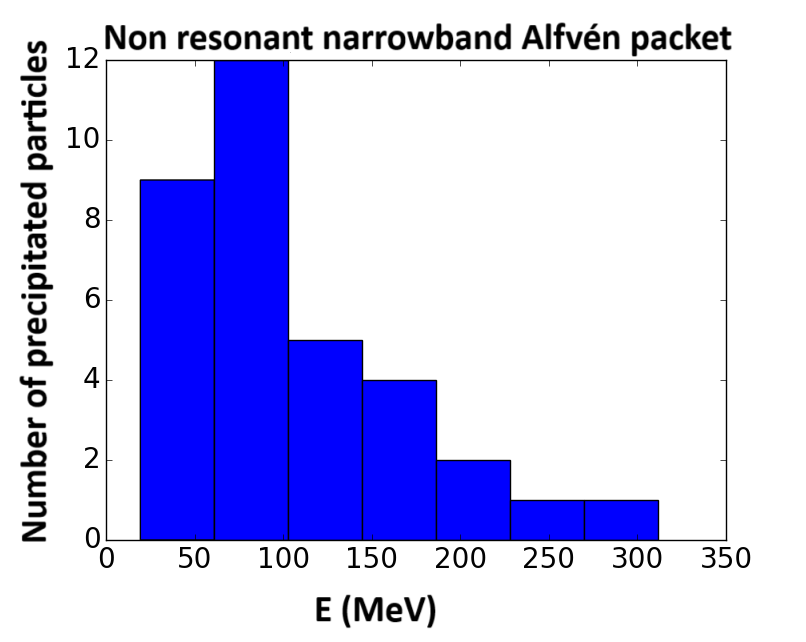}
  \vspace{1em}
 \caption{Energy spectrum of particles lost due to non-resonant Alfvén wave interaction at 40 Hz. The number of precipitated particles is 41, and few particles exceed 100 MeV.}
  \label{fig:energyspectrum_nonresonance}
\end{figure}

For comparison with the resonant case, the energy spectrum of of particles precipitated at a non-resonant interaction frequency of 40 Hz is presented in Fig.~\ref{fig:energyspectrum_nonresonance}.

In this non-resonant case, only 41 particles are precipitated, significantly fewer than in the resonant case. The energy spectrum shows minimal particle precipitation above 100 MeV, confirming that resonant interaction is the dominant mechanism for high-energy particle loss.

 \subsubsection{Broadband Alfv\'en noise}

\begin{figure}[h]
\centering
  \includegraphics[width=0.8\linewidth]{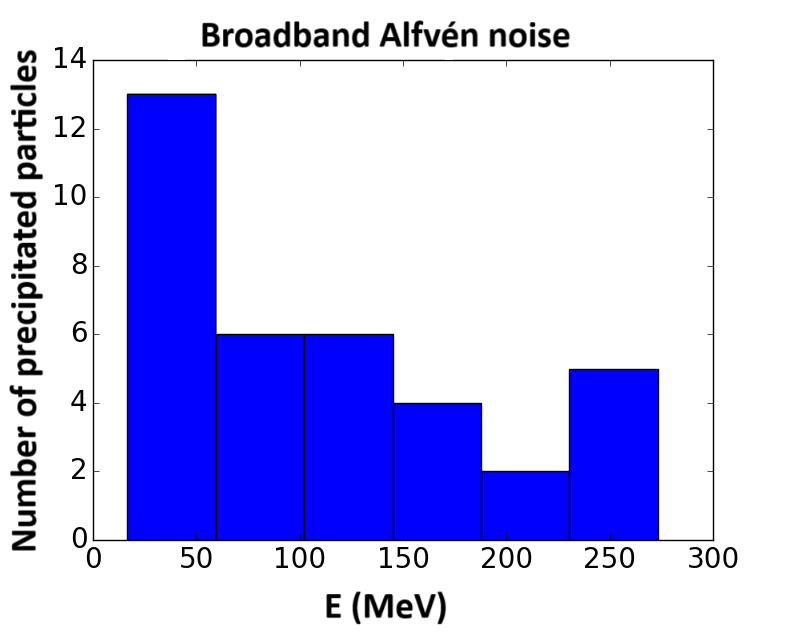}
  \vspace{1em}
 \caption{Energy spectrum of particles precipitated due to interaction with broadband Alfv\'en noise. The spectrum is similar to that of non-resonant narrowband interactions, with minimal loss of high-energy particles.}
  \label{fig:energyspectrum_noise}
\end{figure}

 The energy distribution function $f(E)$, of  particles lost after interaction with background broadband Alfv\'en noise is presented in Fig.~\ref{fig:energyspectrum_noise}.

%This study can be used to distinguish a resonant event in the presence of any background noise.

The results indicate that background broadband Alfvén noise produces only minor particle loss, with the energy spectrum largely similar to non-resonant interactions. This analysis can be used to distinguish a resonant Alfvén event from the background noise.

\textcolor{black}{ We also discuss the mission-level distinguishability of the resonant 10 Hz event from the non-resonant/background case here. In the present simulations, the total particle loss is significantly larger in the resonant 10 Hz case (133 particles) than in the non-resonant 40 Hz case (34 particles) and in the broadband background case (41 particles). Additionally, the resonant narrowband interaction produces a concentrated precipitation peak in the 100–130 MeV range, centered near $\sim 125$ MeV, whereas the non-resonant 40 Hz case and the broadband background produce substantially fewer precipitated particles at comparable energies and exhibit only weak, diffuse precipitation with a relatively flat spectral distribution. Thus, discrimination between resonant and non-resonant 125 MeV events arises primarily from the enhanced count level (higher signal-to-noise ratio) at the resonant energy, together with the presence of a localized spectral peak for the resonant interaction in contrast to the diffuse background of the non-resonant cases. We provide an estimate of the SNR next.}

\textcolor{black}{A simple physics-level signal-to-noise ratio (SNR) may be estimated by defining the signal as the number of precipitated particles in the resonant energy window around the peak and the background as the number of particles in the same energy window for the non-resonant case. From Fig.~17, the resonant case yields approximately $N_{\mathrm{res}} \approx 45$ particles in the 100--125 MeV range, while Fig.~18 shows $N_{\mathrm{bg}} \approx 5$ particles in the corresponding energy range for the non-resonant background. Using Poisson counting statistics, the SNR may be estimated as
\[
\mathrm{SNR} = \frac{N_{\mathrm{res}} - N_{\mathrm{bg}}}{\sqrt{N_{\mathrm{bg}}}} \approx \frac{45-5}{\sqrt{5}} \approx 18,
\]
which indicates that the resonant peak is clearly distinguishable from the non-resonant background.}

\textcolor{black}{We note that a strict detector-level SNR would require folding the simulated spectra through the instrument response and binning them at the 5 MeV energy resolution of the SPEED payload. However, since the predicted peak width ($\sim 30$ MeV) is much larger than the detector resolution, and the estimated counting-statistics SNR is significantly greater than unity, the resonant $\sim 125$ MeV precipitation feature should be resolvable above the non-resonant background.}

\subsection{Count of energetic particle precipitation with altitude}

\begin{figure}[h]
\includegraphics[width=\columnwidth]{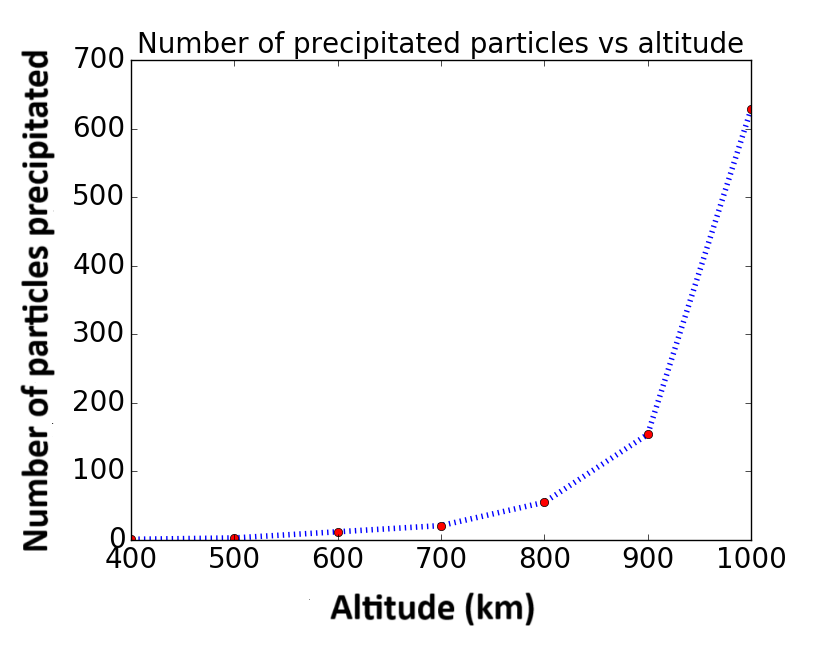}
\centering
\caption{Number of particles detected at different altitudes above the ground. The satellite could be placed in the range (750-850) kms. A satellite placed in the 750–850 km range can observe significant particle precipitation due to wave–particle interactions. At higher altitudes, the satellite may be exposed to high-energy particles, which could be damaging, since the ionosphere–magnetosphere boundary shifts over time, occasionally exposing the satellite to the full flux of energetic particles.}\label{fig:pp_alt}
\centering
\end{figure}

The energetic particle precipitation at different altitudes above the Earth's surface is presented in Fig.~\ref{fig:pp_alt}, \textcolor{black}{which shows the total number of precipitated particles detected at lower altitudes (800 km, 700 km, 600 km, etc.) by progressively lowering the loss-cone (mirror point) precipitation boundary for the same resonant 10 Hz Alfv'en wave. In these simulations, the wave–particle interaction is allowed to continue down to the chosen lowered precipitation boundary, rather than being terminated at 1000 km.}

Our results indicate that the number of energetic particles undergoing precipitation decreases with decreasing altitude. As the precipitation boundary is lowered, fewer particles reach these lower altitudes, indicating that only a reduced fraction of particles is able to penetrate deeper into the atmosphere.

\begin{figure} [h]
\centering
\includegraphics[width=\columnwidth]{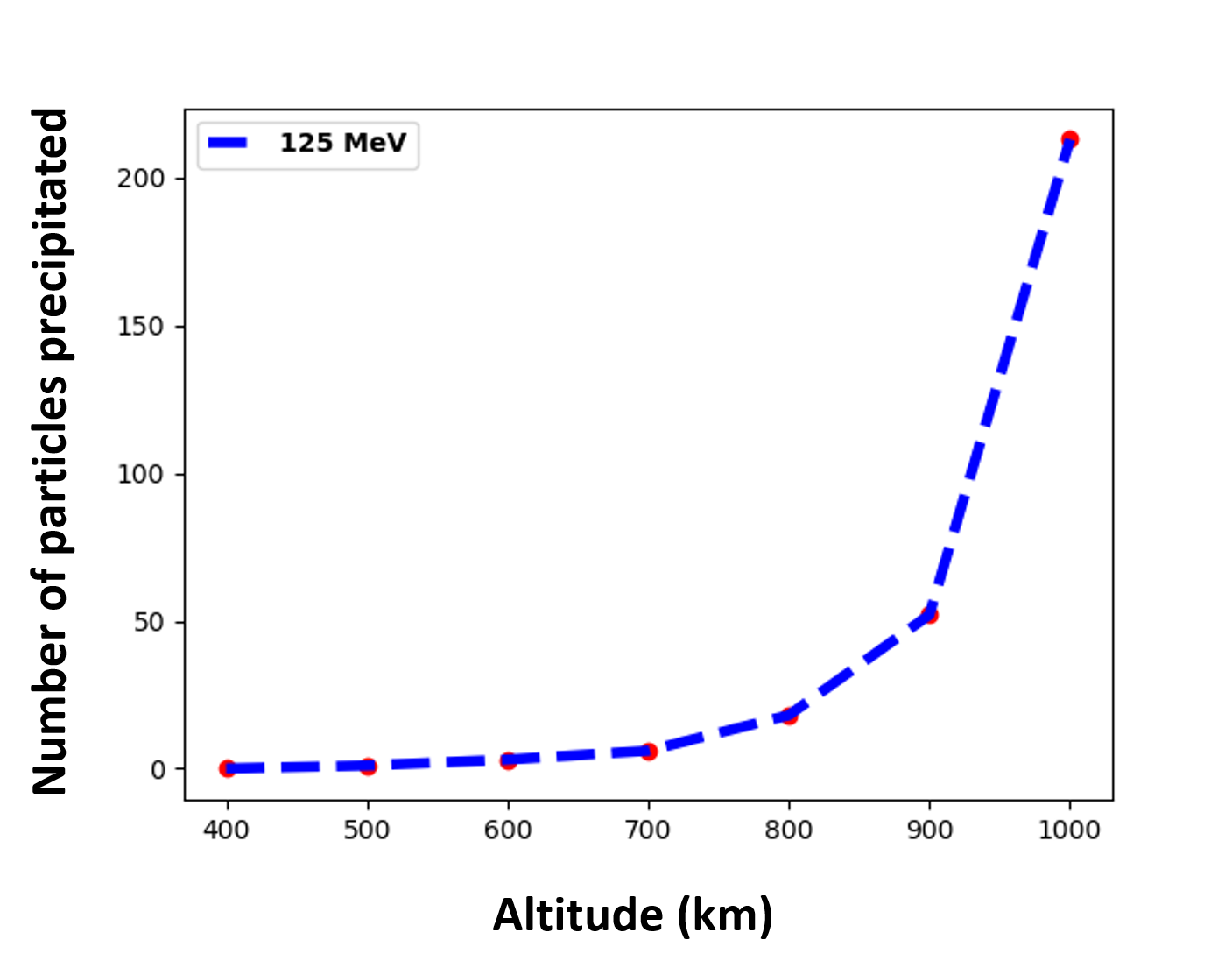}  
  \vspace{1em}
 \caption{This figure shows the number of precipitated particles resulting from resonant wave–particle interaction at different altitudes relevant to the IITMSAT mission, evaluated in the energy band around the 125 MeV resonance peak (100–125 MeV). In the numerical simulations, these altitudes represent the lowered loss-cone (mirror point) precipitation boundary, where particles are considered lost upon reaching the specified altitude. Wave–particle interactions are allowed to continue down to these lowered altitudes, rather than being restricted to the baseline precipitation boundary (1000 km). The plotted counts therefore represent the number of particles that actually reach these altitudes after undergoing wave–particle interaction.}
  \label{125}
\end{figure}

\textcolor{black}{We additionally estimate the number of precipitated particles in the energy band around the 125 MeV resonance peak (100–125 MeV) at different altitudes (see Fig.~\ref{125}).}

\textcolor{black}{We find that the number of peak-energy particles also decreases with decreasing altitude. However, the counts remain substantial in the 800–900 km range, indicating that this altitude band still provides a significant flux of resonant ( $\sim$125 MeV) precipitated protons while sampling particles that have penetrated deeper than those detected at 1000 km. These results suggest that, to capture a significant number of precipitating particles, a satellite should ideally operate within the altitude range of 750–850 km.}

\textcolor{black} {At altitudes closer to 1000 km, the satellite may sample both precipitating particles and the background trapped radiation-belt population, leading to a higher and more continuous flux that reduces the contrast of burst-like signatures. At lower altitudes such as $\sim$800 km, the background trapped population is reduced and the detected flux is dominated by precipitating particles, resulting in clearer intermittent burst signatures and improved detectability.} At higher altitudes, the satellite is also more exposed to energetic radiation-belt particles, increasing the potential risk of radiation damage. In addition, fluctuations in the ionosphere–magnetosphere boundary can intermittently expose the satellite to near-full radiation-belt particle fluxes, further contributing to this risk.

\textcolor{black}{Fig.~\ref{fig:pp_alt} shows that the number of particles precipitated at resonance varies with altitude, with the resonant wave–particle interaction producing different counts of particles reaching different altitudes.  In all these simulations, corresponding to different precipitation boundaries, the energy spectrum of the precipitating particles — specifically the characteristic energy range around the resonant 125 MeV population — remains similar (see Fig.~\ref{fig:boat1}). Thus, although the particle counts differ significantly, the energy spectra of precipitating particles at different altitudes are broadly similar. The resonance condition primarily determines the energy of the precipitating particles and therefore preserves a similar spectral shape across altitudes.}

\section{Summary}

In this paper, we have presented results from numerical simulations carried out to study the interactions between steady state energetic trapped protons and low-frequency (ULF) Alfvén wave packets in the inner Van Allen radiation belt. Initially, we developed a 3D kinetic model of the inner radiation belt, which was constructed to approximately match the observed particle density. We then performed FDTD simulations of MHD Alfvén wave packets and studied their interactions with the steady-state trapped particle population in the belt. The main results of our study are summarized below:

\begin{enumerate}

%\item An $\alpha = 10 $ represents the best observed density for particles in the inner radiation belt. The analytical method gives  approximate particle guiding centre trajectory whereas the numerical simulation takes care of true gyro-motion. Hence, the loss cone could be properly studied using the current numerical simulations.

\item An $\alpha = 10$ provides the best match to the observed particle density in the inner radiation belt. While the analytical model approximates particle guiding-center trajectories, the numerical simulations capture full gyro-motion, allowing a more accurate study of the loss cone and particle precipitation.

\item A numerical study of the interactions of narrow-band Alfv\'en wave packets with the energetic particles in the belt shows a sharp cyclotron resonance condition between low Alfv\'en frequencies around 10 Hz and high energy protons around 125 MeV due to which there is substantial particle  precipitation at lower altitudes from their stable mirror orbits.

\item This resonant event is clearly distinguished from other non resonant interactions or background broadband Alfv\'en noise, which does not cause significant energetic particle precipitation. This distinction has been confirmed from our numerical simulations.

\item To detect significant energetic particle precipitation due to these bursts, a satellite should ideally orbit at ~800 km altitude. \textcolor{black}{At higher altitudes ($\sim$1000 km), the satellite samples both precipitating and trapped particles, reducing burst contrast, whereas at $\sim$800 km the flux is dominated by precipitating particles, yielding clearer signatures. Higher altitudes also entail increased radiation exposure due to enhanced background flux and boundary fluctuations, raising the risk of damage.} \textcolor{black}{The energy spectra of resonant precipitating particles at different altitudes are broadly similar, as the resonance condition primarily determines the characteristic energy. }

\end{enumerate}

\textcolor{black}{We also address the issue of distinguishing a ``seismic'' particle burst from a ``geomagnetic'' one, and examine whether specific spectral signatures in precipitated protons can be associated with 10~Hz Alfv\'enic interactions, as considered in the present study of seismic-induced bursts. We note that the present model focuses on precipitation driven by seismic forcing, and geomagnetic-storm--related processes are not explicitly included. Therefore, a direct discrimination between seismic and geomagnetic particle bursts is beyond the scope of this study.}

\textcolor{black}{Seismo-electromagnetic emissions (SEME) are proposed to generate ULF disturbances that can couple to MHD Alfv\'en waves in the inner radiation belt. In this framework, we investigate narrowband ULF Alfv\'en wave--energetic proton interactions and find that an Alfv\'enic wave centered at 10~Hz produces resonance-driven pitch-angle scattering and precipitation of protons around $\sim 125$~MeV, consistent with the cyclotron resonance condition. The resulting particle burst in the model is therefore interpreted as ``seismic,'' arising from a frequency-selective resonant interaction.}

\textcolor{black}{In contrast, observational studies of geomagnetic-storm proton dynamics in the inner radiation belt\citep{Zhang2021}, such as Zhang \textit{et al.} (2021, \textit{Chinese Phys. B} \textbf{30}, 129401), indicate that storm-time variations are broadband and governed by multiple mechanisms. Low-energy protons (2--20~MeV) are primarily affected by Coulomb collisions associated with enhanced atmospheric density, intermediate energies (30--100~MeV) are influenced by magnetic field-line curvature scattering, and higher-energy protons ($>100$~MeV) are largely
unaffected during storms. These processes therefore produce energy-dependent, broadband flux variations rather than a narrow, resonance-driven burst.}

\textcolor{black}{The present simulations instead exhibit a distinct resonance-controlled signature. When scanning Alfv\'en wave frequencies in the range 5--50~Hz, maximum precipitation is obtained near 10~Hz, corresponding to cyclotron resonance for $\sim 125$~MeV protons. The resulting precipitation is localized in both energy and pitch-angle space, with the particle spectrum concentrated predominantly in the range 100--130~MeV and peaking near 125~MeV. This narrowband response, tied to a specific wave frequency, contrasts with the broadband, mechanism-dependent nature of geomagnetic-storm--driven losses. Notably, the 100--130~MeV proton range is reported to be relatively unaffected during geomagnetic storms, and ULF activity associated with storms has not been shown to produce comparable resonance-driven proton loss in the inner belt.}

\textcolor{black}{Therefore, within the present modeling framework, a seismic particle burst may therefore be characterized by a narrowband, resonance-driven precipitation peak near 10~Hz and a localized proton energy signature around $\sim 125$~MeV, whereas geomagnetic-driven precipitation is expected to exhibit broader, energy-dependent variations without a dominant spectral peak. This distinction suggests a potential spectral diagnostic for identifying precipitation associated with 10~Hz Alfv\'enic interactions, although further observational validation is required.}

\textcolor{black}{In the present model, energetic protons are treated as test particles interacting with prescribed Alfvén waves, without accounting for their collective feedback on the wave dynamics. In a fully self-consistent collisionless plasma, the energetic proton distribution could give rise to collective effects which can enhance or suppress energetic particle precipitation into the loss cone. For example, energetic protons can interact with Alfvén waves, producing collective effects — including modifications of wave amplitudes and spectra through resonant interactions (linear/quasilinear effects) and nonlinear processes that occur when waves become sufficiently strong — which can influence precipitation into the loss cone. Energetic protons with an anisotropic distribution can drive or damp Alfvén waves, similar to anisotropy-driven instabilities, enhancing or reducing wave amplitudes — which in turn increases or decreases pitch-angle scattering, thereby modifying precipitation into the loss cone.  Resonant interactions  at specific frequencies can also alter the wave spectrum, changing the amplitude versus frequency profile of the waves and consequently affecting precipitation characteristics. 
Nonlinear effects arise when waves become strong enough to trap particles in their potentials, leading to phenomena such as proton trapping and phase-space bunching, the formation of coherent structures, and other nonlinear processes including frequency shifts, wave–wave coupling, and nonlinear Landau damping — all of which can influence precipitation characteristics.
These collective effects, resulting from interactions of the energetic protons with the fields they collectively generate, are not included in the present study and remain a topic for future investigation.}

\bibliography{CR}
\bibliographystyle{ieeetr} % Cite with full title.
%\bibliographystyle{apsrev} % Cite without title.
%\clearpage
\appendix
\setcounter{figure}{0}
\renewcommand{\thefigure}{A\arabic{figure}}

%\clearpage

{
\color{black}
\section{RESONANT WAVE-PARTICLE INTERACTION}\label{AppenB}

\setcounter{figure}{0}
\renewcommand{\thefigure}{B\arabic{figure}}

\color{black}
\subsection{Diffusion coefficient from resonant wave-particle interaction and comparison with Quasi-Linear Theory}\label{AppenB1}

In the present manuscript, the resonant interaction is illustrated by the 125 MeV proton interacting with the 10 Hz Alfv\'en wave. For this case, we can make an order-of-magnitude comparison between the diffusion implied by the simulation and the expectation from quasi-linear theory (QLT).

From Fig.~20, the magnetic moment of the resonant 125 MeV particle changes by an amount of order
\[
\Delta \mu \sim 10^{-7}
\]
during the interaction. Since the particle--wave interaction in the simulation is evolved over a time interval of
\[
\Delta t \sim 1\,{\rm s},
\]
a finite-time estimate of the magnetic-moment diffusion coefficient ($D_{\mu\mu}$) is
\[
D_{\mu\mu}^{\rm sim}
\sim \frac{(\Delta\mu)^2}{2\Delta t}.
\]
Substituting the above values gives
\[
D_{\mu\mu}^{\rm sim}
\sim \frac{(10^{-7})^2}{2 \times 1}
\sim 5 \times 10^{-15},
\]
in units of $\mu^2\,{\rm s^{-1}}$.

To convert this to the pitch-angle diffusion coefficient ($D_{\alpha\alpha}$), we use the relation between magnetic moment ($\mu$) and pitch angle ($\alpha$),
\[
\mu = \frac{m v^2}{2B}\sin^2\alpha.
\]
Differentiating with respect to $\alpha$ gives
\[
\frac{d\mu}{d\alpha}
= \frac{m v^2}{B}\sin\alpha \cos\alpha.
\]
Since
\[
\mu = \frac{m v^2}{2B}\sin^2\alpha,
\]
this may also be written as
\[
\frac{d\mu}{d\alpha} = 2\mu \cot\alpha.
\]

We first estimate the particle pitch angle from the representative resonant-particle velocities already used in the resonance-condition check (see Sec.~4.2.1, “Energy Spectrum and Pitch-Angle Scattering of Precipitated Particles,” under “Narrowband Alfvén wave packet in calculation of Cyclotron Resonance Condition”). With
\[
v_\perp \approx 1.30 \times 10^8\,{\rm m\,s^{-1}}, 
\qquad
v_\parallel \approx 8.94 \times 10^7\,{\rm m\,s^{-1}},
\]
the corresponding pitch angle is
\[
\alpha = \tan^{-1}\!\left(\frac{v_\perp}{v_\parallel}\right)
       = \tan^{-1}\!\left(\frac{1.30 \times 10^8}{8.94 \times 10^7}\right)
       \approx 55.5^\circ.
\]

Using the representative resonant-particle values
\[
\mu \approx 0.85 \times 10^{-6},
\qquad
\alpha \approx 55.5^\circ,
\]
we obtain
\[
\frac{d\mu}{d\alpha}
= 2 \mu \cot\alpha
\approx 2(0.85 \times 10^{-6}) \cot(55.5^\circ)
\approx 1.17 \times 10^{-6}.
\]

The pitch-angle diffusion coefficient is then related to the magnetic-moment diffusion coefficient by
\[
D_{\alpha\alpha}^{\rm sim}
\approx
\frac{D_{\mu\mu}^{\rm sim}}
{\left( d\mu/d\alpha \right)^2}.
\]
Substituting the above values gives
\[
D_{\alpha\alpha}^{\rm sim}
\sim
\frac{5 \times 10^{-15}}
     {(1.17 \times 10^{-6})^2}
\sim
3.7 \times 10^{-3}\,{\rm s^{-1}}.
\]

We now compare this with a quasi-linear estimate. For weak resonant Alfv\'enic scattering, an order-of-magnitude quasi-linear scaling for pitch-angle diffusion is
\[
D_{\alpha\alpha}^{\rm QLT}
\sim
\frac{\pi}{2}\,\Omega_c
\left(\frac{\delta B}{B_0}\right)^2
\sin^2\alpha.
\]
For the representative resonant particle, we use
\[
\alpha \approx 55.5^\circ,
\qquad
\sin^2\alpha \approx 0.68,
\]
\[
\Omega_c = \frac{qB}{m}
\approx 1.60 \times 10^3\,{\rm s^{-1}}
\quad
\text{for } B = 16\,\mu{\rm T},
\]
and from the wave-amplitude estimate in the manuscript,
\[
\frac{\delta B}{B_0} \sim 4 \times 10^{-4} \; \text{to} \; 10^{-3}.
\]

Using these values, we obtain
\[
D_{\alpha\alpha}^{\rm QLT}
\sim
\frac{\pi}{2}
(1.60 \times 10^3)
\left(4 \times 10^{-4}\text{--}10^{-3}\right)^2
(0.68),
\]
which gives
\[
D_{\alpha\alpha}^{\rm QLT}
\sim
2.7 \times 10^{-4}
\; \text{to} \;
1.7 \times 10^{-3}\,{\rm s^{-1}}.
\]

For completeness, the corresponding quasi-linear estimate of the magnetic-moment diffusion coefficient is
\[
D_{\mu\mu}^{\rm QLT}
\approx
\left(\frac{d\mu}{d\alpha}\right)^2
D_{\alpha\alpha}^{\rm QLT}.
\]
Using
\[
\left(\frac{d\mu}{d\alpha}\right)^2
\approx
(1.17 \times 10^{-6})^2
\approx
1.37 \times 10^{-12},
\]
we obtain
\[
D_{\mu\mu}^{\rm QLT}
\sim
(1.37 \times 10^{-12})
\left(2.7 \times 10^{-4}\text{--}1.7 \times 10^{-3}\right),
\]
i.e.
\[
D_{\mu\mu}^{\rm QLT}
\sim
3.7 \times 10^{-16}
\; \text{to} \;
2.3 \times 10^{-15},
\]
again in units of $\mu^2\,{\rm s^{-1}}$.

Thus, for the resonant 125 MeV proton interacting with the 10 Hz Alfv\'en wave, the finite-time simulation-based estimate gives
\[
D_{\alpha\alpha}^{\rm sim} \sim 3.7 \times 10^{-3}\,{\rm s^{-1}},
\]
while the quasi-linear estimate gives
\[
D_{\alpha\alpha}^{\rm QLT} \sim 10^{-4}\text{--}10^{-3}\,{\rm s^{-1}}.
\]
Similarly,
\[
D_{\mu\mu}^{\rm sim} \sim 5 \times 10^{-15},
\qquad
D_{\mu\mu}^{\rm QLT} \sim 10^{-16}\text{--}10^{-15}.
\]

We therefore find that the diffusion implied by the simulation is broadly consistent, to within an order of magnitude, with the quasi-linear expectation for weak resonant scattering. We emphasize that this comparison is intended as an order-of-magnitude consistency check rather than an exact equality, since the simulation estimate is based on a finite-time change of a single strongly interacting particle, whereas quasi-linear theory describes an ensemble-averaged diffusion process.

\subsection{"Few hour" Lead time: Basis?}\label{AppenB1}

The “few hours” lead time mentioned in the manuscript primarily refers to the observational interval between the energetic particle burst and the occurrence of the main earthquake. This interval largely reflects the earthquake preparation process, during which stress accumulation, microfracturing, crack growth, and charge activation/separation develop prior to the final rupture. The crust does not fail instantaneously; rather, the source region evolves toward instability, and the associated electromagnetic signatures may appear before the main shock, giving rise to pre-seismic disturbances.

The wave propagation time from the source region (hypocenter) through the Earth's crust to the surface, and subsequently through the atmosphere to the ionosphere/magnetosphere  (e.g., ULF electromagnetic propagation through the atmosphere and along geomagnetic field lines), is expected to be much shorter—typically of the order of seconds to minutes. A rough calculation is provided next.

Since the crust is a conducting medium, the disturbance propagates diffusively with a characteristic time $\tau_{\mathrm{diff}} \sim d^{2}/\eta = \mu_{0}\sigma d^{2}$; taking $d \sim 30$ km, $\mu_{0}=4\pi\times10^{-7}\ \mathrm{H,m^{-1}}$, and a representative crustal conductivity $\sigma \sim 10^{-3}$–$10^{-1}\ \mathrm{S,m^{-1}}$ gives $\tau_{\mathrm{diff}} \sim (4\pi\times10^{-7})(10^{-3}\text{--}10^{-1})(3\times10^{4})^{2} \approx 1$–$100$ s, indicating that electromagnetic disturbances generated at hypocentral depths can reach the surface on timescales of order seconds to minutes. From the surface to the ionosphere at altitude $h \sim 100$ km, the electromagnetic propagation time is  $\tau_{\mathrm{atm}} \sim h/c \sim 100/(3\times10^{5})\ \mathrm{s} \approx 3\times10^{-4}$ s,  so this stage is essentially instantaneous compared to the other timescales.  Once coupled to the magnetic field, the disturbance propagates along the field line over a path length  $s \sim 1.2\times10^{4}$ km for the $L=1.5$ shell, with Alfv\'en speed 
$V_A \sim (1.5$--$8.5)\times10^{3}\ \mathrm{km\,s^{-1}}$. Using the lowest Alfv\'en speed gives the largest propagation time,
\[
\tau_A \sim \frac{s}{V_A} \sim \frac{1.2\times10^{4}}{1.5\times10^{3}} \ \mathrm{s} \approx 8\ \mathrm{s}.
\]

Adding these contributions, the total propagation time can be written as
\[
\tau_{\mathrm{total}} \sim \tau_{\mathrm{crust}} + \tau_{\mathrm{atm}} + \tau_A
\approx (1\text{--}100)\ \mathrm{s} + \text{negligible} + 8\ \mathrm{s}
\sim 10^{2}\ \mathrm{s},
\]
i.e., of the order of a few minutes, which is much shorter than the observed few-hour lead time.

 The particle bounce time along L shell 1.5 in our case is only $\sim 0.2$–0.25 s, as estimated and also obtained from the numerical simulations. 
 
The time required for finite pitch-angle diffusion can be estimated using
\[
\tau_{\mathrm{diff}} \sim \frac{(\Delta\alpha)^2}{D_{\alpha\alpha}} .
\]
From the simulation, the change in magnetic moment is $\Delta\mu \sim 10^{-7}$, and using 
\[
\frac{d\mu}{d\alpha} \approx 1.17\times10^{-6},
\]
gives a representative pitch-angle change
\[
\Delta\alpha \sim \frac{\Delta\mu}{d\mu/d\alpha}
\approx \frac{10^{-7}}{1.17\times10^{-6}}
\approx 0.085\ {\rm rad}.
\]
Using the simulation-based diffusion coefficient 
\[
D_{\alpha\alpha}^{\rm sim} \sim 3.7\times10^{-3}\ {\rm s^{-1}},
\]
we obtain
\[
\tau_{\mathrm{diff}} \sim \frac{(0.085)^2}{3.7\times10^{-3}}
\approx \frac{7.2\times10^{-3}}{3.7\times10^{-3}}
\approx 2\ {\rm s}.
\]

Thus, pitch-angle diffusion in the resonant interaction occurs on timescales of seconds to at most a minute, which is again much shorter than the observed few-hour lead time. 

The few-hour lead time therefore reflects the temporal evolution of pre-seismic lithospheric processes (stress accumulation, microfracturing, and charge activation), whereas wave propagation, wave–particle interaction, and pitch-angle diffusion occur on much shorter timescales of seconds to minutes. In many cases, observational studies have reported associated ionospheric and energetic particle anomalies from a few hours to a few days before some earthquakes.

\end{document}